\documentclass[aps,prd,twocolumn,groupedaddress,showpacs,tighten]{revtex4}
\usepackage{times}
\usepackage{epic,epsfig,axodraw}
\usepackage{subfigure}

\def\be{\begin{equation}}
\def\ee{\end{equation}}
\def\bea{\begin{eqnarray}}
\def\eea{\end{eqnarray}}
\def\ba{\begin{array}}
\def\ea{\end{array}}
\def\ghost#1{}

\def\beq{\begin{equation}}
\def\eeq{\end{equation}}
\def\bey{\begin{eqnarray}}
\def\eey{\end{eqnarray}}

\def\lsim{\mathrel{\raise.3ex\hbox{$<$\kern-.75em\lower1ex\hbox{$\sim$}}}}
\def\gsim{\mathrel{\raise.3ex\hbox{$>$\kern-.75em\lower1ex\hbox{$\sim$}}}}

\begin{document}

\title{\vspace*{2mm}
\boldmath $U$-boson production in $\,e^+e^-$ annihilations, $\,\psi\,$ and 
$\,\Upsilon$ decays, \,and Light Dark Matter\\
\vspace{3mm}}
\author{Pierre Fayet$^{1}$\\ \vspace{7mm}}

\affiliation{$^1$Laboratoire de Physique Th\'eorique de l'ENS, UMR 8549 CNRS, 24 rue Lhomond, 75231 Paris Cedex 05, France\vspace{2mm}}

\date{\vspace*{9mm} February 17, 2007}

\begin{abstract}
{\vspace{1mm}
We recall how a new light gauge boson emerged in Supersymmetric extensions of the Standard Model with an extra singlet chiral superfield, and how it could often behave very much as a
light pseudoscalar, with the corresponding symmetry broken at a scale higher than electroweak.
\vspace{1mm}\\ \indent
The possible existence of such a new gauge boson $U$, light and very weakly coupled, 
allows for Light Dark Matter particles, which could be at the origin of the 511 keV line 
from the galactic bulge. 
Could such a light gauge boson 
be found directly in $e^+e^-$ annihilations\,?
\,Not so easily, in fact, due to various constraints 
limiting the size of its couplings, 
especially the axial ones, leading to an axionlike behavior or extra parity-violation effects.
In particular, searches for the decay $\ \Upsilon\to\gamma\,+ $ invisible $U\,$
may be used to constrain severely the {\it \,axial\,} coupling of the $U$ to the electron, 
$\,f_{e\,A}=f_{b\,A}$, \,to be less than about $ \,10^{-6}\ m_U$\,(MeV), \,50 times smaller than the 
$\simeq \,5\ 10^{-5}\ m_U$\,(MeV) \,that could otherwise have been allowed from $\,g_e-2$.
 \vspace{1mm}\\ \indent
The {\it \,vector} coupling of the $U$ to the electron may in principle be larger, but is also limited in size. Even under favorable circumstances
(no axial couplings to quarks and charged leptons, and very small couplings to neutrinos), 
taking also into account possible $Z$-$U$ mixing effects,
we find from $\,g_\mu-2\,$, under reasonable assumptions (no cancellation effect, 
lepton universality), 
\,that the vector coupling of the $U$ to the electron can be at most as large as $ \,\simeq 1.3\ 10^{-3}$,
\,for $\,m_U<m_\mu\,$. \,Such a coupling to the muon of the order of $\,10^{-3}$
could also be responsible 
for the somewhat large value of the measured $\,g_\mu-2\,$, \,as compared to Standard Model expectations, should this effect turn out to be real.
\vspace{1mm}\\\indent
The $U$ couplings to electrons are otherwise likely to be smaller, \,e.g. $\lsim 3\ 10^{-6}\ m_U$\,(MeV), 
if the couplings to neutrinos and electrons are similar.
This restricts significantly the possibility of detecting a light $\,U$ boson 
in $\ e^+e^-\to\gamma\,U$, \,making this search quite challenging.  
Despite the smallness of these couplings, 
$\,U$ exchanges can provide annihilation cross sections of LDM particles 
of the appropriate size, even if this may require that light dark matter 
be relatively strongly self-interacting.
\\
}
\end{abstract}
\pacs{\ 12.60.Cn, 13.20.Gd, 13.66.Hk, 14.70.Pw, 
95.35.+d
\hspace{20mm} LPTENS-07/07}


\maketitle

Theories beyond the Standard Model often involve extended gauge groups, 
neces\-sitating new spin-1 gauge bosons,
in addition to the gluons, photon, $W^\pm$ and $Z$.
It is usually believed that they
should be heavy 
\,($\gsim\, $ several hundred GeV's at least) or even very heavy, as in grand-unified theories,
in which they could mediate proton decay.
Still some could be light, even very light, 
provided they are, also, very weakly coupled, and therefore neutral.

\section{A \,light $\, U$ boson}
\label{sec:1}

We discussed, long ago, the possible existence of such a new gauge boson called $\,U$,
exploring in particular
limits on its production and decay (depending on its mass)
into $\,e^+e^-$ or $\,\nu \bar\nu\,$ pairs ...~\cite{fayet:1980rr}.
Such a particle originated from supersymmetric extensions of the Standard Model, 
which require {\it two} electroweak doublet Higgs superfields,
offering the possibility, in non-minimal versions of the Supersymmetric Standard Model 
with an extra chiral singlet superfield \cite{susy,nmssm}
of ``rotating'' independently the two doublets, i.e. of gauging an extra-$U(1)$ symmetry.
The standard gauge group is then extended to $\,SU(3)\times SU(2)\times U(1)\ \times$ extra-$U(1)$.

\vspace{2mm}

The fact that the effects of such a gauge boson did not show up
in neutrino-scattering experiments
(and a possible connection of this spin-1 particle with gravity through the massive spin-$\frac{3}{2}$ gravitino 
\cite{grav})
led us to consider that it could be both {\it \,light\,} and {\it \,very weakly coupled}.
Its mass is generated through the v.e.v.'s of the two Higgs doublets $h_1$ and $h_2$,
plus a possible singlet, of the supersymmetric extensions 
of the Standard Model. Or also, in a similar way, in non-supersymmetric extensions as well, 
in which case a single Higgs doublet, plus an additional singlet, may be sufficient.

\vspace{2mm}
The phenomenology of a light neutral 
spin-1 $U$ boson, independently of its possible origin,
turns out to be quite rich.
It could be produced in $ \,q \bar q \,$ or $\,e^+ e^-\,$ annihilations through processes like
\be
\label{prodU}
\psi \,\to\, \gamma\ U\,, \ \ \ \Upsilon \,\to\, \gamma\ U\,, \ \ \ K^+\,\to\,\ \pi^+\,U\ \ ,
\ee
\vspace{-5mm}

\noindent
and
\vspace{-4mm}
\be
e^+e^-\, \to \,\gamma\ U\ \ ,
\ee
including even positronium decays, should the $U$ be lighter than 1 MeV 
(cf. Figs.\,\ref{fig:eegammau} and {\ref{fig:upsilon} in Sections \ref{sec:6} and 
\ref{sec:11})
\cite{fayet:1980rr,fayetmez}.
It could also lead to interesting effects in neutral-current phenomenology, 
including neutrino scatterings, anomalous magnetic moments of charged leptons, 
parity-violation in atomic physics, ...
(cf. Figs.~\ref{fig:magn},\,\ref{fig:limite},\,\ref{fig:pvat},\,\ref{fig:neut}
\,in Sections \ref{sec:8},\,\ref{sec:9},\,\ref{sec:11},
\ref{sec:13}) \,\cite{fayet:1980rr,pvat,pvat2,pvat3}.

\vspace{2mm}

The $U$ boson could also be extremely light (or maybe even massless), with extremely small couplings (down to $\,\approx 10^{-19}$ and less).
Its vector couplings are normally expected, for ordinary neutral matter, to be expressed as a linear combination of the conserved (or almost conserved) $B$ and $L$ currents, 
or $\,B-L\,$ in a grand-unified theory (rather than to other quantities like strangeness or mass)
\cite{equiv}.
It could then lead to apparent violations of the equivalence principle;
and in the massive case to possible deviations from the $\,1/r^2$ law of gravity, 
the new force induced by $U$ exchanges having a finite range $\,\hbar/(m_U c)$ \cite{cras}.
Both effects have been searched for experimentally, and are constrained by \cite{adel}.
But this is not a situation we shall be interested in here, as we shall consider 
much larger values of the $U$ mass \,-- more than 1 MeV --\, and of the gauge couplings of the $U$ boson
to quarks and leptons, typically $\,\gsim 10^{-6}\,$.

\vspace{2mm}

We shall mainly be interested in the direct production of a $U$ boson in the process $e^+e^-\to\gamma\,U$, discussing what magnitude may be expected 
for its scattering cross section, given
that \hbox{$U$-induced} annihilations, represented in Fig.~\ref{fig:ann} of Section \ref{sec:5},
may  also be responsible for an appropriate relic density
of {\it \,light dark matter}  (LDM) particles ~\cite{boehmfayet, fermion}, 
which could be at the origin of the 511 keV line from the galactic bulge 
\cite{integral, boehm511}.

\vspace{2mm}
Estimating this cross section requires taking into account a variety of constraints, 
especially those involving {\it \,axial couplings\,} of the $\,U$
(from $\,\psi\,$ and $\,\Upsilon\,$ decays, $\,g_\mu-2$, \,and parity-violation in atomic physics), \,as well as the fact that the $\,U$
should in general couple to the electroweak Higgs doublet(s) and therefore mix with the $Z$.
We shall also see that LDM annihilations do not really constrain significantly the size of the $\,U$ couplings to the electron. But other processes severely limit them, and therefore the detectability of the direct production of a $\,U$ boson in $\ e^+e^-\to\,\gamma\ U\,$.

\section{Enhanced \,effects \,of the \,axial \,couplings 
\vspace{1mm} \break
of a \,light $\,U\!$}
\label{sec:2}

\vspace{-1mm}

\noindent
{\it ``Axionlike'' behavior of a light $\,U$} \cite{fayet:1980rr} {\it \,:}

\vspace{2mm}

If the gauge couplings $f_{q,l}$ of the new spin-1 boson $U$ with quarks and leptons are very small, 
it looks like the $U$ should be very weakly coupled to these particles, almost by definition.

\vspace{2mm}
This is, however, not necessarily true\,! How is it possible\,?
Even with such very small couplings, the rates for producing a light $\,U$ through its interactions with quarks and leptons, 
although seemingly proportional to $\,f_{q,l}^{\ 2}\,$,
would not necessarily be small in the presence of axial couplings.

\vspace{2mm}

Indeed a non-vanishing axial coupling \,($f_{q,l \ A}$)\, of the $\,U$ to a quark or lepton
would generate, 
for a longitudinally-polarized $\,U$ (with $\,\epsilon^\mu_{\,L}\simeq \,k^\mu/m_U$), 
\,an effective pseudoscalar coupling
\be
\label{fqlp}
f_{q,l\ p}\ \,=\ \,\frac{2\ m_{q,l}}{m_U}\ \ f_{q,l \ A\,}\ \ .
\ee
This one may be sizeable, even if the axial gauge coupling $f_{q,l \ A\,}$ is very small, if the mass of the $U$ boson is small as well.
In fact, this axial coupling $f_{q,l \ A}$ simply regenerates in a spontaneously broken gauge theory, through eq.~(\ref{fqlp}), 
the pseudoscalar couplings to quarks and leptons of the spin-0 Goldstone boson 
\,(denoted by $a$)\,
that was eliminated when the $U$ acquired its mass.
\,A light \hbox{spin-1} $U$ boson would then be produced, through its interactions with quarks and leptons,
like this spin-0 pseudoscalar (i.e.~also very much like a 
\hbox{spin-0} axion),
 proportionally to $\,f_{q,l\ A}^{\ 2}/m_U^{\,2},$
\,times $\,m_{q,l}^{\ 2}\,$, \,i.e. to $\,f_{q,l\ p}^{\ 2}\,$.

\vspace{4mm}

\noindent
{\it Supersymmetry spontaneously broken ``at a high scale'':}

\vspace{2mm}

In a similar way the $\pm\frac{1}{2}$ polarisation states of a massive but very light \hbox{spin-$\frac{3}{2}$} gravitino, 
although coupled only with extremely small gravitational
strength \,(i.e. proportionally to $\kappa = \sqrt{8\pi\,G_{\rm Newton}}\,
\simeq \,4\,10^{-19}\  \hbox{GeV}^{-1}$), 
would undergo enhanced gravitational interactions, owing to the large factor
\be
\sqrt{\,\frac{2}{3}\,}\ \ \ \frac{k^\mu}{m_{3/2}}\ \ 
\ee
then present in the expression of the gravitino wave function \cite{grav}.
Although still coupled with gravitational strength $\,\propto\kappa\,$, \,these states would be produced 
and interact much more strongly, proportionally to $(\kappa^2/m_{3/2}^{\,2})\ ...$\,, with 
the gravitino mass $m_{3/2}$ expressed as 
\be
m_{3/2}\ =\ \kappa d/\sqrt 6\ ,\ \ \ \hbox{or}\ \ \ \kappa F/\sqrt 3\ \ .
\ee
These interaction or decay rates involving 
light gravitinos are proportional to $\,\kappa^2/m_{3/2}^{\,2}\,$ i.e. to $\,1/d^2\,$ or $\,1/F^2$,
\linebreak
\,where 
$\sqrt d\,/2^{1/4}=\sqrt F=\Lambda_{\rm ss}$ is usually called the supersym\-metry-breaking scale, 
so that
\be
\Lambda_{\rm ss}\,=\,(3/8\pi)^{1/4}\,\sqrt{\,m_{3/2}\,m_P}\ \ .
\ee

The $\pm\frac{1}{2}$ polarisation states of a light gravitino would behave, in fact, very much like a spin-$\frac{1}{2}$ goldstino \cite{grav}. The strength of these enhanced gravitino interactions, fixed by the gravitino mass $m_{3/2}$ or equivalently the supersymmetry-breaking scale, could be sizeable if supersymmetry were broken ``at a low scale'', comparable to the electroweak scale, the gravitino mass 
being then very small
(e.g.~\,typically $\,\propto\,\hbox{(electroweak scale)}^2/m_{\rm Planck}\approx 10^{-5}$~eV$/c^2$).
But this strength would become very small, or again extremely small 
\,-- with the corresponding spin-$\frac{1}{2}$ goldstino state very weakly or extremely weakly coupled --\,
if supersymmetry gets broken ``at a large scale''.
The gravitino then acquires a sizeable mass
\be
m_{3/2}\ \,=\,\ \sqrt{\,\frac{8\pi}{3}}\  \ \frac{\Lambda_{ss}^2}{m_P}\ \ ,
\ee
possibly up to 
$\,\sim m_W$ to TeV scale, supersymmetry being then said to be broken at the
scale $\,\Lambda_{\rm ss} \sim\,10^{10}\,$ to $ \,10^{11}$ GeV
\footnote{The process $\,e^+e^-\to\gamma\,U$, 
\,with an amplitude proportional to the extra-$U(1)$ 
gauge coupling of the $U$ to the electron,
may also be partly related with the corresponding process $\,e^+e^-\to$ $ \tilde\gamma\,\tilde G\,$,
$\,\tilde G\,$ denoting the massive spin-$\frac{3}{2}\,$ gravitino (or the equivalent spin-$\frac{1}{2}$ goldstino) \cite{grav}, with an amplitude inversely proportional to the supersymmetry-breaking scale parameter $d$. This happens 
if the $U$ contributes in part (through the v.e.v. 
of the corresponding auxiliary $D$ component) to the generation of mass splittings between bosons and fermions within the multiplets of supersymmetry.
{\it \,(Very) small gauge couplings} of the $U$ to quarks and leptons would then 
correspond to a $\sqrt{\hbox{\small$\,<\!D\!>$}}\,$ (and therefore $\,\Lambda_{\rm ss}$) \
{\it (much) higher than electroweak scale}.}.

\pagebreak

\noindent
{\it ``Hiding'' these enhanced effects of axial couplings, with an extra-$U(1)$ symmetry broken 
at a higher scale:}

\vspace{1.5mm}

Let us return to spin-1 particles, with very small gauge couplings to quarks and leptons.
The smallness of the couplings of a massive gauge particle
is not sufficient to guarantee that its interactions will actually also be small
\,(as we saw above for a \hbox{spin-$\frac{3}{2}$} particle),
\,if this spin-1 particle has non-vanishing axial couplings.
This requires, in fact, that the scale at which the corresponding (extra-$U(1)$) symmetry 
is spontaneously broken be sufficiently large \,(as for a massive gravitino and supersymmetry-breaking scale, in supersymmetric theories).

\vspace{2mm}

Searches for such light $U$ bosons with non-vanishing axial couplings, 
as in the hadronic decays (\ref{prodU}) of the $\,\psi$, $\,\Upsilon$, \,or $\,K^+$,
with the $U$ decaying into unobserved $\nu\bar\nu$ or light dark matter particle pairs,
then require, dealing with standard model particles,
that the extra-$U(1)$ symmetry be broken at a scale higher 
than the electroweak scale.
And possibly even at a large scale if an extra singlet acquires a large vacuum expectation value, possibly much higher than the electroweak scale,
according to a mechanism already exhibited in \cite{fayet:1980rr}
and which also applies to spin-0 axions as well, making them ``invisible''.

\section{\vspace{1mm} Gauging \,and \,breaking \,the \,extra-$U(1)_A$ symmetry} \label{sec:3}

In the absence of such an extra singlet, a light spin-1 $U$ boson would behave very much like a light spin-0 pseudoscalar $A$ described by a linear combination of the neutral Higgs doublet components $\,h_1^{\,\circ}$ and $h_2^{\,\circ}$, reminiscent of a standard axion, or of the $A$ of the MSSM when this one is light.

\vspace{3mm}

\noindent
{\it Two Higgs doublets and their v.e.v.'s:}

\vspace{1.5mm}
Let us denote
\be
h_1\ =\ 
\hbox{\small$\displaystyle
\left(\ba{c} 
h_1^{\,\circ} \vspace{1mm}\\ h_1^{\,-} 
\ea \!\right)
$}\ , \ \ \
h_2\ =\ 
\hbox{\small$\displaystyle
\left(\ba{c} 
h_2^{\,+} \vspace{1mm}\\ h_2^{\,\circ}  
\ea \!\right)
$}\ \ ,
\ee
the two Englert-Brout-Higgs doublets whose v.e.v.'s
\be
\label{higgsvev}
<\!h_1^{\,\circ}\!>\ =\frac{v_1}{\sqrt 2}\ =\frac{v}{\sqrt 2}\, \cos \beta\,,\ \ \,
<\!h_2^{\,\circ}\!>\ =\frac{v_2}{\sqrt 2}\ =\frac{v}{\sqrt 2}\, \sin \beta\,,
\ee
are responsible for the masses of down quarks and charged leptons, and up quarks, respectively, as in supersymmetric extensions of the Standard Model  \,-- although one may also choose not to work within 
supersymmetry, or disregard the SUSY sector of $R$-odd superpartners.
We denote
\be
\frac{1}{x}\ \,=\,\ \tan\beta\ \,=\,\ \frac{v_2}{v_1}\ \ ,
\ee
which replaces the $\,\tan\delta=v'/v"\,$ of ~\cite{susy,nmssm}, with
\be
\varphi"=
\hbox{\small$\displaystyle
\left(\ba{c} 
\varphi"^\circ \vspace{1mm}\\ \varphi"^- 
\ea \!\right)
$}
\to \,h_1 \ , \ \ 
\varphi'=
\hbox{\small$\displaystyle
\left(\ba{c} 
\varphi'^\circ \vspace{1mm}\\ \varphi'^- 
\ea \!\right)
$}
\ \ \hbox{with}\ \ \
\varphi^c\ \to \,h_2\ .
\ee

\pagebreak

\noindent
{\it Gauging an $U(1)_A\,$:}

\vspace{2mm}

Of course in a supersymmetric theory there is here no $\,\mu\,\,H_1 H_2$ superpotential term as it would not be invariant under the extra-$U(1)$ symmetry that we intend to gauge, if one is to rotate independently the two Higgs doublets $h_1$ and $h_2$, using as in \cite{fayet74}
the invariance under
\be
\label{uh1h2}
h_1\ \to\ e^{i\,\alpha}\ h_1\ \,,\ \ \ h_2\ \to\ e^{i\,\alpha}\ h_2\ \,,
\ee
and similarly for the 
two doublet Higgs superfields $H_1$ and $H_2$.

\vspace{2mm}

The $\,\mu\,$ parameter was in fact promoted to a full chiral superfield in \cite{nmssm}, the 
$\,\mu\,\,H_1 H_2$ term being replaced by a trilinear coupling with 
an extra singlet chiral superfield $N$  \footnote{The $\,\mu\,$ parameter was often considered, later, as a source of difficulty referred to as the \,``$\mu$~problem''. 
This could be taken as one further reason for getting rid of the $\,\mu\,$ term in favor of a trilinear coupling $\,\lambda\ H_1 H_2\,N$. 
~However, the size of the supersymmetric $\,\mu\,$ parameter may be controlled by considering 
either an approximate extra-$U(1)$ symmetry such as the one we gauge here, 
or an approximate continuous $R$-symmetry (broken, at the energy scale of SUSY particles, by the gravitino and gaugino mass terms, in particular), 
as $\,\mu\,$  occurs in violation of both symmetries. 
This allows one to evacuate the so-called $\,\mu$ problem, 
without necessarily having to replace the $\,\mu$ term by a trilinear coupling with the singlet $N$.
},
\be
\mu \ H_1 H_2\ \ \to\ \ \lambda\ \,H_1 H_2 \,N\ \ .
\ee 
This replacement of the $\,\mu\,$ term by a trilinear $\ \lambda\ H_1 H_2 \,N$ coupling allowed, subsequently, 
for the gauging \cite{susy} of an extra-$U(1)\,$ symmetry acting as in (\ref{uh1h2}), already identified in \cite{nmssm} under the name of $\,U$, \,under which
\be
\label{Usym}
H_{1,2}\ \to\ e^{i\,\alpha}\ H_{1,2}\ ,\ \ N\ \to\ e^{-\,2\,i\,\alpha}\ N\ \ ,
\ee
so that $\ \lambda\ H_1H_2\,N\,$ is $\,U$-invariant, but not $N$ itself \footnote{As 
the linear term $\,\sigma N\,$ used in the superpotential of \cite{nmssm}
explicitly broke this (then ungauged) extra-$U(1)$ symmetry,
our two Higgs doublets generated, with the ``$R$-invariant'' superpotential
$$
\lambda\ \,H_1H_2\,N\ +\ \sigma\ N
$$
({\it not\,} invariant under any other extra-$U(1)$ symmetry),
 a spontaneous breaking of $SU(2)\times U(1)$ down to $\,U(1)_{\rm QED}$, 
\,without unwanted massless or quasimassless 
(``axionlike'' or ``dilatonlike'') spin-0 particles.}.

\vspace{2mm}

The gauging of this extra-$U(1)\,$ symmetry \footnote{This gauging was also motivated by other reasons,
like, at the time, making squarks and sleptons heavy through sponta\-neously-generated $D$ terms 
rather than using (universal) explicit dimension-2 $\,m_\circ^{\,2}\,$ squark and slepton mass$^2$ terms, 
as already introduced in the first paper of \,\cite{susy} 
but breaking explicitly the supersymmetry. 
Generating spontaneously these $\,m_\circ^{\,2}\,$ terms led us to gauge an extra-$U(1)_A$ symmetry acting {\it axially} (in the simplest case)
on quarks and leptons.
$m_\circ^{\,2}$ terms are now usually generated through gravity-induced breaking.
}, 
in the presence of the $\,\lambda\ H_1 H_2\,N\,$ trilinear superpotential coupling,
therefore requires not to include in the superpotential
any of the $\,N,\,N^2$ and $\,N^3$ terms \cite{susy}.
(Of course we do not have to gauge such an extra-$U(1)$ symmetry, in which case we remain 
with one version or the other \,-- depending on which of the $\,N,\ N^2\,$ and $\,N^3\,$ terms 
are selected in the $N$ superpotential \footnote{The last two terms, proportional to $\,N^2$ and $\,N^3$, are also forbiden by the continuous $R$-symmetry if it is imposed \cite{nmssm}. This one gets reduced to $R$-parity in the presence of gravity, owing to the gravitino and gaugino mass terms, in particular \cite{grav}.
It is still possible to use $R$-symmetry to forbid the $N^3$ term
(corresponding to dimension-4 terms in the Lagrangian density), 
while allowing for gravity-induced terms of dimensions $\leq 3$ associated 
with $\,m_{3/2}$, \,for which $R$-symmetry is reduced to $R$-parity.}
--\, of a non-minimal $\,SU(3)\times SU(2)\times U(1)\,$ supersymmetric extension of the Standard Model, often called the NMSSM \cite{susy,nmssm}.)

\vspace{2mm}
In any case, this construction allows for the generation of quark and charged-lepton masses,
in a way compatible with the gauging of the extra-$U(1)$ symmetry,
through the usual trilinear superpotential
\be
\lambda_e\ H_1\ L\,\bar E\ +\ \lambda_d\ H_1\ Q\,\bar D\ -\ \lambda_u\ H_2\ Q\,\bar U\ \ ,
\ee
leading from (\ref{higgsvev}) to charged-lepton and quark masses
\be
m_e\ =\ \lambda_e\ \frac{v_1}{\sqrt 2}\ \ ,\ \ \ 
m_d\ =\ \lambda_d\ \frac{v_1}{\sqrt 2}\ \ ,\ \ \ 
m_u\ =\ \lambda_u\ \frac{v_2}{\sqrt 2}\ \ ,
\ee
$SU(2)$ and family indices being omitted for simplicity.

\vspace{2mm}

This extra-$U(1)$ symmetry acts in the simplest case on the left-handed (anti)quark and (anti)lepton
superfields as follows \cite{susy}
\be
(Q,\,\bar U,\, \bar D;\ L,\,\bar E)\ \ \to\ \ e^{-\,i\frac{\alpha}{2}}\ (Q,\,\bar U,\, \bar D;\ L,\,\bar E)\ \ ;
\ee
i.e. it acts {\it axially\,}  on quark and lepton fields,
\be
\label{uql}
\left\{
\ba{cccc}
\hbox{\em doublets:} &(q_L,\ l_L) & \ \to \  & e^{-\,i\frac{\alpha}{2}}\ \ (q_L,\ l_L)\ \ ,
\vspace{3mm}\\
\hbox{\em singlets:}& (u_R,\ d_R,\ e_R) & \ \to\  & e^{i\frac{\alpha}{2}}\ \ (u_R,\ d_R,\ e_R)\ \ ,
\ea \right.
\ee
with family indices again omitted for simplicity, together with
\be
h_1\ \to\ e^{i\,\alpha}\ h_1\ ,\ \ \ h_2\ \to\ e^{i\,\alpha}\ h_2\ \ ,
\ee
as in (\ref{uh1h2},\ref{Usym}).

\vspace{3.5mm}

\noindent
{\it 
The Goldstone boson of $U(1)_A$, \,and the axion:}

\vspace{1.5mm}
This extra-$U(1)_A$ symmetry acts on quarks, leptons and the two Higgs doublets 
as the one considered in \cite{pq} in connection with
the strong $CP$ problem. The corresponding Goldstone boson $\,a\,$ considered here 
is eaten away when the extra $U(1)$ is gauged so that the corresponding gauge boson acquires a mass.
Constructed from the two neutral Higgs doublet components $h_1^{\,\circ}\,$ and $\,h_2^{\,\circ}\,$ (plus a possible singlet contribution as we saw) \cite{fayet:1980rr,susy}, this would-be masless Goldstone boson 
$\,a$ is reminiscent of a spin-0 axion \cite{wilc,weinberg}.

\vspace{2mm}
The $\,U(1)$ of \cite{pq}, however, is intrinsically anomalous and corresponds
to a pseudo-symmetry violated by quantum effects, 
to ``rotate away'' the $CP$-violating parameter $\theta\,$ of QCD,
the corresponding pseudo-Goldstone boson, called axion, acquiring a small mass.
The extra-$U(1)$ symmetry should here, in principle, be made anomaly-free if it is to be gauged, even if the cancellation of anomalies may involve a new sector of the theory, not necessarily closely connected to the one discussed here.
The spin-0 Goldstone boson $a$ gets eliminated when the spin-1 $U$ boson acquires its mass.

\vspace{2mm}

\noindent
{\it Cancelling anomalies:}

\vspace{1mm}

The extra-$U(1)$ symmetry discussed above would be anomalous,
if we limit ourselves to the quarks and leptons of the standard model.
Anomalies may be cancelled,
e.g.~by extending the theory to include new mirror quarks and leptons ($q^m$ and $l^m$), transforming 
under the extra $U(1)$ as follows:
\be
\left\{
\ba{cccc}
\hbox{\em doublets:} & (q_R^m,\ l_R^m)& \ \to\ & \ e^{-\,i\frac{\alpha}{2}}\ \ (q_R^m,\ l_R^m)\ \ ,
\vspace{3mm}\\
\hbox{\em singlets:}  &(u_L^m,\ d_L^m,\ e_L^m) & \ \to \ & 
\ e^{i\frac{\alpha}{2}}\ \ (u_L^m,\ d_L^m,\ e_L^m)\ \ , \vspace{1mm} \\
\ea \right.
\ee
the counterpart of (\ref{uql}), 
so that the whole theory be vectorlike.

\vspace{2mm}

Since $\ \overline{d_L^m}\,q_R^m\,$ transforms like $\,\overline{d_R}\,q_L\,$
under $\,SU(2)\times U(1)\ \times$ extra-$U(1)\,$, etc.,
$\,<\!h_1^{\,\circ}\!>\,$ and $\,<\!h_2^{\,\circ}\!>\,$ may 
(just as for ordinary quarks and leptons) 
be responsible for mirror charged-lepton and down-quark masses, 
and mirror up-quark masses, respectively 
(through Yukawa couplings proportional to 
$\ h_1\ \overline{d_L^m}\,q_R^m\,+$ h.c.\,, \,etc.),
in a two-Higgs-doublet theory, 
in a way compatible with the extra-$U(1)$ symmetry 
--\, but ignoring for the moment supersymmetry.

\vspace{2mm}

In a supersymmetric theory however, we have to take into account the analyticity of the superpotential. $H_1$ and $H_2$ may still be used to generate mirror quark and lepton masses
through superpotential terms proportional to
$H_2\ \bar L^m E^m, \ H_2\ \bar Q^m D^m$ and $\,H_1\ \bar Q^m U^m$, \,in a $SU(3)\times SU(2)\times U(1)$ gauge theory,
but this cannot be done in a way compatible with the above extra-$U(1)$ symmetry. 
Indeed, as the mirror (anti)quark and (anti)lepton superfields (still taken left-handed)
transform as follows:
\be
(\bar Q^m,U^m,D^m;\bar L^m,E^m)\, \to\, 
e^{i\frac{\alpha}{2}}\ (\bar Q^m,U^m,D^m;\bar L^m,E^m)\ ,
\ee
we need to introduce two more doublet Higgs superfields, $H_3$ and $H_4$ (again taken as left-handed, with opposite weak hypercharges $Y=\pm1$) transforming under the extra $U(1)$ according to
\be
H_{3,4} \ \,\to\,\   e^{-i\alpha}\, \ H_{3,4}\ \ ,
\ee
so as to generate mirror quark and lepton masses in an extra-$U(1)$-invariant way \cite{n=2}.

\vspace{2mm}
They appear in fact as the mirror counterparts of $\,H_1\,$ and \,$H_2$, \,also required to avoid anomalies associated with the extra-$U(1)$ couplings of the two higgsino doublets $\,\tilde h_1\,$ and $\,\tilde h_2$ (cf. eq.~(\ref{Usym})), so that the whole theory be vectorlike.
This is also reminiscent of $\,N\!=\!2\,$ extended supersymmetric theories, which naturally involve
({before $N\!=\!2\,$ supersymmetry breaking)
{\it \,four doublet Higgs superfields\,} rather than the usual two,
then describing, in particular, 4 Dirac charginos, etc. \cite{2guts}.

\ghost{
\footnote{These two additional Higgs doublets are required in addition to give masses 
to mirror quarks and leptons in a way invariant under the extra-$U(1)$ symmetry as we explained,
as well as to avoid anomalies associated with the extra-$U(1)$ couplings of the two higgsinos
described by $H_1$ and $H_2$, the extra two Higgs superfields are also required,
within $N=2$, to give masses to all \,-- i.e. now four --\,  Dirac charginos.
}.
}

\vspace{3mm}
Instead of gauging the extra $U(1)$ as discussed here, one may also consider a global (and possibly anomalous) \,extra-$U(1)$ symmetry
spontaneously or explicitly broken (e.g.~by $N, \, N^2$ or $N^3$ superpotential terms, 
or soft supersymmetry-breaking terms). It then generates a massless Goldstone boson $a$, or a would-be (pseudo-)Goldstone boson, which acquires a mass (small if the amount of explicit breaking 
of the extra $U(1)$ is small.

\vspace{2mm}
In all these cases, the branching ratios for $\,\psi$ \,(or $\,\Upsilon\,$)\,
$\to\,$ light spin-1 $U$ boson, or light spin-0 pseudoscalar $a$, will be essentially the same.
Let us now discuss the couplings to quarks and leptons of the spin-1 $\,U\,$ boson, or of its ``equivalent'' spin-0 pseudo\-scalar~$\,a$.

\section{\vspace{1mm} 
Couplings of the equivalent \vspace{1mm} \break
spin-0 \,pseudoscalar $\, a$}
\label{sec:3bis}

\vspace{-1mm}

The Yukawa couplings of the two Higgs doublets $\,h_1$ and $\,h_2$ to  quarks and leptons are
\be
\lambda_{d,l}\, = \, \frac{m_{d,l}}{v_1/\sqrt2}\ = \ \frac{m_{d,l}}{\frac{v}{\sqrt2}\, \cos\beta}\ ,\ \ 
\lambda_u\, =\, \frac{m_{u}}{v_2/\sqrt2}\ =\ \frac{m_{u}}{\frac{v}{\sqrt2} \,\sin\beta}\ ,
\ee 
and those of their real neutral components 
($\,\sqrt 2\  \Re\,h_1^{\,\circ}$\, and 
\,$\sqrt 2\   \Re\,h_2^{\,\circ}$)\,, 
\be
\label{couph1h2}
\left\{\ 
\ba{ccc}
\hbox{\small $\displaystyle \frac{m_{d,l}}{v_1}$} &=&
\displaystyle 2^{1/4}\ G_F^{1/2} \ \,m_{d,l}\,/\,\cos\beta \ \ , 
\vspace{3mm}\\
\hbox{\small $\displaystyle \frac{m_u}{v_2}$}  &=&
\displaystyle 2^{1/4}\ G_F^{1/2}\, \ m_{u}\,/\,\sin\beta\ \ , 
\ea \right.
\ee 
respectively \footnote{
The couplings of the SM Higgs field \,($\sqrt 2\ \Re\,\varphi^\circ$)\,
are given, in terms of $\ <\!\sqrt 2\,\ \Re\,\varphi^\circ\!> \ \,=\,v \simeq 246$ GeV,
\,by $\,\frac{m_{q,l}}{v}=2^{1/4}\,G_F^{1/2}\ m_{q,l}\,$.
}.
As $\,m_t/m_b = (\lambda_t/\lambda_b) \times \,(v_2/v_1)$, 
\,larger values of $\,1/x=\tan\beta\ $ 
(between $\,\approx 1\,$ up to $\,\approx \,m_t/m_b\,\simeq 40$) \,may be preferred.

\vspace{2mm}

The massless Goldstone boson field eliminated away by the massive $Z$,
previously denoted in \cite{susy,nmssm} as $\ \sqrt 2\ \ \Im\,(\cos\delta \ \varphi"^\circ +\sin\delta\ \varphi'^\circ)\,$,
\,reads in modern notations
\be
z_g\ =\ \sqrt 2\ \ \Im\ (\cos\beta \ h_1^{\,\circ}-\sin\beta\ h_2^{\,\circ})\ \ .
\ee
Its orthogonal combination
\be
\label{A}
A\ \ =\ \ \sqrt 2\ \ \Im\ (\sin\beta \ h_1^{\,\circ}\,+\,\cos\beta\ h_2^{\,\circ})\ \ 
\ee
(ignoring for the moment possible extra singlet v.e.v.'s) represents, in the presence of the new extra-$U(1)$ symmetry, the massless spin-0 Goldstone field to be eliminated by the $U$, when the $\ SU(3)\times SU(2)\times U(1)_Y\times\ $ extra-$U(1)\,$ symmetry gets spontaneously broken down to $SU(3)_{\rm QCD}\times U(1)_{\rm QED}\,$ through $\,<\!h_1^{\,\circ}\!>\,$ 
and $\,<\!h_2^{\,\circ}\!>\ $
\footnote{Within supersymmetric theories, these two fields $z_g\,$ (Goldstone boson eliminated by the $Z$) and $A$ (pseudoscalar to be eliminated later by the $U$)
are described by the two orthogonal chiral superfield combinations 
$\,H_g= (\cos\beta \ H_1^{\,\circ}-\sin\beta\ H_2^{\,\circ})\,$ and
$\,(\sin\beta \ H_1^{\,\circ}+\cos\beta\ H_2^{\,\circ})\,$, \,respectively.}
\footnote{This $A$ field, that became in \cite{susy} a massless Goldstone boson eliminated away when the extra-$U(1)$ is gauged so that the \hbox{spin-1} $U$ boson acquires a mass, was formerly massive in \cite{nmssm}, as the superpotential used there,
$\,\lambda\ H_1H_2N+\sigma\,N$, breaks explicitly this extra-$U(1)$ symmetry, 
so that the existence of a massless or quasimassless axionlike pseudoscalar was automatically avoided.\\
On the other hand, if we consider $\,\lambda=0\,$ (e.g. by taking the limit in which 
$\lambda\,$ and $\,\sigma\,$ get small, their ratio, and therefore $v_1 v_2$ being fixed), we return to a situation in which the extra-$U(1)$ is spontaneously broken,
$\,A$ being the corresponding Golsdtone boson, also associated with a massless or quasimassless \hbox{spin-0} scalar (``modulus'') corresponding to another flat direction of the potential, 
as for $\,\lambda=0\,$ the minimisation of  $\,(\vec D^2+D'^2)/2\,$ in $V$ 
only determines $\ |v_2|^2-|v_1|^2\,$.
\,Both bosons are described by $\,(\sin\delta\ \varphi''^\circ+\cos\delta\ {\varphi'^\circ}^*)$
\,(eq.~(53) of \cite{nmssm}),
\,i.e. in modern notations $\,(\sin\beta\ h_1^\circ+\cos\beta\ h_2^\circ)$, 
the spin-0 component of 
$\,(\sin\beta\ H_1^\circ+\cos\beta\ H_2^\circ)$\,.}.

\vspace{2mm}

With $h_1$ and $h_2$ separately responsible for down-quark and charged-lepton masses, and up-quark masses, 
respectively, as in supersymmetric theories,
we get from (\ref{couph1h2},\ref{A}) the usual expression of the pseudoscalar couplings of
$\,A\,$ to quarks and charged leptons,
\bea
\label{couplagep}
\left\{
\ba{cl}
2^{1/4}\,G_F^{1/2}\ m_{u}\ \cot\beta \ \ \ [\hbox{or}\ x]\,, &\hbox{\small  for $u$-quarks},
\vspace{3mm}\\
2^{1/4}\,G_F^{1/2}\ m_{d,l}\ \tan\beta \ \ [\hbox{or}\ \ 
\hbox{\small$\displaystyle\frac{1}{x}$}]\,,&
\hbox{\small for $d$-quarks and ch. leptons},
\ea \right.
\nonumber
\eea
\vspace{-8mm}
\be
\ee
which acquire their masses through $\,<h_2^{\,0}>\,$ and  $\,<h_1^{\,0}>$, \,respectively\,\footnote{If we forget about supersymmetry 
we might decide that $u$-quarks\,/\,$d$-quarks\,/\,charged-leptons 
get masses indifferently from couplings to either $\,h_1$, or $\,h_2$. The resulting pseudoscalar 
couplings of $A$ would then be $\,2^{1/4}\,G_F^{1/2}\ m_{q,l}\,$ times $\,x\,$ 
for the fermions acquiring masses through $\,<\!h_2\!>\,$ (ordinarily up quarks); or $\,1/x\,$ for those acquiring masses through $\,<\!h_1\!>\,$ 
(ordinarily down quarks and charged leptons).
This analysis applies as well to such situations.
If two different doublets are separately responsible for all quark masses
(say $h_2$)
and charged-lepton masses (say $h_1$),  the limits from $\,\psi$ and $\,\Upsilon$ 
decays would no longer directly restrict the size of the axial couplings to charged leptons,
$\,f_{e\,A}$.}.

\vspace{2mm}

In the presence of one or several extra singlets transforming under the extra-$U(1)$ symmetry and acquiring non-vanishing v.e.v.'s \cite{fayet:1980rr}, expression (\ref{A}) of the equivalent spin-0 pseudoscalar
gets modified, to
\be
\label{expressiondea}
a\ =\ \cos\zeta \ \ \hbox{(\,``standard'' $A$\,)}
\vspace{2mm}\\
+\ \sin\zeta \ \
\hbox{(\,new singlet\,)}\,,
\ee
in which we define
\be
r\ \,=\,\ \cos\zeta\ \ .
\ee
The spin-1 $U$ boson, instead of behaving like the \hbox{spin-0} pseudoscalar $A$
given by (\ref{A}),
i.e. very much like a standard axion, 
now behaves (excepted for the $\,\gamma\gamma\,$ coupling, absent) like the above doublet-singlet combination $\,a$.

\vspace{2mm}

As the extra spin-0 singlets are not directly coupled to quarks and leptons,
the effective pseudoscalar couplings of $U$ to quarks and charged leptons read
\be
\label{couplagep2u}
\ba{ccl}
2^{1/4}\ G_F^{1/2}\ \ m_{u}\ \ \ r\, x\,
&\simeq& \, 4\ 10^{-6}\ m_u\,\hbox{(MeV)}\ \ \ r\, x\ \vspace{2mm}\\
&\simeq& \, 4\ 10^{-6}\ m_u\,\hbox{(MeV)}\ \ \cos\zeta \, \cot\beta \vspace{1mm}\\
\ea 
\ee
for up quarks, and 
\be
\label{couplagep2d}
\ba{ccl}
2^{1/4}\ G_F^{1/2}\ m_{d,l}\ \ \ \,\hbox{\small$\displaystyle\frac{r}{x}$}\,\ 
&\simeq&\,  4\ 10^{-6}\ m_{d,l}\,\hbox{(MeV)}\ \ \ \,\hbox{\small$\displaystyle\frac{r}{x}$}
\vspace{2mm}\\
&\simeq& \, 4\ 10^{-6}\ m_u\,\hbox{(MeV)}\ \ \cos\zeta \, \tan\beta\ 
\vspace{1mm}
\\
\ea 
\ee
for down quarks and charged leptons.

\vspace{2mm}

The $\,\psi\to\gamma\,U$ and $\Upsilon\to\gamma\,U$ decay rates, in particular, 
are multiplied by the factor
\be
r^2\ =\ \cos^2\zeta\ <\ 1\ \ ,
\ee
which may be small.
This corresponds precisely to the mechanism by which the standard axion 
may be replaced by a new axion, called later ``invisible''.
As for such an axion, all amplitudes for emitting or absorbing (resp. exchanging) in this way 
a light $U$ boson are multiplied by the parameter $\,r=\cos\zeta\leq 1$ (resp. $r^2\leq 1$), 
\,which becomes very small 
when the extra singlet acquires a large v.e.v. \cite{fayet:1980rr}.

\vspace{2mm}

The corresponding axial couplings of the $U$, in general obtained after taking into account $Z$-$U$ mixing effects (cf. next Section), are then given by
\be
\label{faxr}
f_{q,l\,A}\, =
\underbrace{\small \ 2^{-3/4}\ G_F^{1/2}\ m_U\ }_{\displaystyle 2\ 10^{-6}\ m_U\hbox{(MeV)}} \!
\times
\left\{
\ba{cl}
\displaystyle r\,x\,, &\!\!\hbox{\small for up quarks}\,,
\vspace{3mm}\\
\hbox{\small$\displaystyle \frac{r}{x}$}\,,&
\!\!\hbox{\small for $d$-quarks and ch. lept.}.
\ea \right.
\ee
in agreement with eq.~(\ref{fqlp}).

\section{Implications \,for  $\ \,e^+e^-\to\gamma \,U\,$:}
\label{sec:4}

\vspace{-3mm}
\begin{center}
{\bf \small a first discussion}
\end{center}

\vspace{-.5mm}

Altogether {\it \,the axial couplings of a $U$ boson $\,f_{q,l\,A}\,$ turn out to be rather strongly constrained}, especially for light $U$, owing to the enhancement factor 
$\,2\,m_{q,l}/m_U\,$ appearing in eq.~(\ref{fqlp}).
We are going to discuss here, in particular, the effects of this phenomenon on the possible size of the couplings of the $U$ boson to the electron.

\vspace{3mm}

\noindent
{\bf \small Constraint on the axial couplings of the \boldmath $\,U\,$  from \boldmath $\,g_\mu-2$\,:}

\vspace{1.5mm}

Let us consider the contribution to the anomalous magnetic moment of the muon,
induced by the exchange of a virtual $U$ boson (see Fig.~\ref{fig:limite} 
in Section \ref{sec:9}).
If the $U$ is significantly lighter than the muon
there is an enhancement of the effects of its axial coupling,
by a factor $\approx \,(4)\ m_\mu^{\,2}/m_U^{\,2}\,$ originating from the
expression of its propagator,
\be
\label{propagator}
\frac{-\ g^{\mu\nu}+\frac{\hbox{$k^\mu\,k^\nu$}}{\hbox{$m_U^{\ 2}$}}}{k^2-m_U^{\ 2}}\ \ \,.
\ee
This enhancement factor, $\,\approx (4)\ 100\,$ for a 10 MeV $U$ boson, 
could lead a too large negative contribution to $\,g_\mu-2$,
\,proportional to $\,f_{\mu\,A}^{\ 2}/m_U^2\,$.
More precisely 
\be
\label{amuaxial}
\delta\,a_\mu^A\ \,\simeq\,\ 
-\ \,\frac{f_{\mu \,A}^{\ 2}}{4\,\pi^2}\ \ \frac{m_\mu^{\,2}}{m_U^{\ 2}}\ \
= \ \,-\ \,\frac{f_{\mu\,p}^{\ 2}}{16\,\pi^2}\ \ 
\ee
is found (owing to (\ref{fqlp})) to be essentially the same as for the exchange of the equivalent 
pseudoscalar spin-0 particle $a$. \,I.e. also the same as for a standard axion, 
times the factor $\,r^2 = \cos^2\zeta \,\leq 1\,$ associated with the fact that an extra Higgs singlet may acquire a (possibly large) v.e.v., increasing the scale at which the extra-$U(1)$ symmetry gets spontaneously broken, as compared to the electroweak scale \cite{fayet:1980rr,pvat}. 

\vspace{2mm}

In agreement with expression (\ref{fqlp}) of the equivalent pseudoscalar coupling, 
\be
\label{fmupseudo}
f_{\mu\ p}\ \,=\ \,\frac{2\ m_\mu}{m_U}\ \ f_{\mu A\,}\ \,=\, \ 2^{1/4}\ \,G_F^{1/2}\ m_\mu\ \frac{r}{x}\ \ ,
\ee
we get for this axial contribution, ``enhanced'' by the effect of the factor $\,m_\mu^{\,2}/m_U^{\,2}\,$
but now also reduced by the extra factor
$\,r^2=\cos^2\zeta\ $ \cite{pvat},
\be
\label{amuaxial2}
\ba{ccl}
\delta\,a_\mu^A\ &\simeq&
\displaystyle
-\ \frac{G_F\,m_\mu^{\,2}}{8\,\pi^2\,\sqrt 2}\ \ \frac{r^2}{x^2}\ 
\simeq\ -\ 1.17\ \,10^{-9}\ \ \frac{r^2}{x^2}\ \ 
\vspace{3mm}\\
&\simeq&\ \ \ \ \ \ -\ 1.17\ \,10^{-9}\ \ \cos^2\zeta\ \,\tan^2\beta\ \ .
\ea
\ee

\vspace{2mm}

In the absence of approximate cancellations with other (positive) contributions, such as those that would be induced by the vector couplings of the $U$, 
$\,\delta a_\mu^V \simeq f_{\mu V}^{\,2}/(8\,\pi^2)\,$ for a sufficiently light $U$,
this leads as in \,\cite{fermion} to
a rather severe constraint on $\,f_{\mu\,p}\,$, $\,r/x\!<1\,$
(cf. Section \ref{sec:9}).
It corresponds, owing to (\ref{faxr}), to
\be
\label{limfmua}
|f_{\mu\,A}|\ \,\lsim\ \,2\ \,10^{-6}\ \,m_U\hbox{(MeV)}\ \ ,
\ee
approximately expressed as
\be
\label{limfmua2}
\frac{f_{\mu\,A}^{\ 2}}{m_U^{\,2}}\ \,\lsim\ \,\frac{G_F}{3}\ \ .
\ee

This constraint on $\,f_{\mu\,A}\,$
\,-- only valid in the absence of cancellations with other positive contributions to $\,\delta a_\mu$ --\,
 may be applied to the axial coupling to the electron, 
under the reasonable hypothesis of lepton universality, also in agreement with eq.~(\ref{faxr}) giving the axial couplings of the $U$ within the class of models considered.
The resulting constraint, i.e. (\ref{limfmua}) and (\ref{limfmua2}})  
now applied to $\,f_{e\,A}$, \,turns out to be significantly more restrictive than the corresponding one,
\be
|f_{e\,A}| \,\lsim\ \,5\ \,10^{-5}\ \,m_U\hbox{(MeV)}\ \ ,
\ee
that follows directly from the $\,g_e\!-2\ $ of the electron
\,(cf. Section \ref{sec:8}).

\vspace{3mm}

\noindent
{\bf \small  Constraints on axial couplings from quarkonium decays\,:}

\vspace{2mm}

The axial couplings of the $U$ to the $c$, $b$ and $s$ quarks get also constrained from
the $\,\psi\to\gamma\,U\,$, $\,\Upsilon\to\gamma\,U\,$ and $\,K^+\to\pi^+\,U\,$ decays, respectively
(cf. Section \ref{sec:11}).
In simple situations ensuring the absence of unwanted flavor-changing neutral current effects
\cite{equiv,U} (cf. Section \ref{sec:10}), 
{ \it \,the axial couplings to the charge $-\frac{1}{3}\,$ $d$, $s$ and $b$ quarks 
are found from gauge invariance to be equal to  $f_{e\,A}$}. 
This is, of course, also in agreement with expression (\ref{faxr}) of the axial couplings of the $U$, 
related through (\ref{fqlp}) to expressions (\ref{couplagep2u},\ref{couplagep2d}) 
of the equivalent pseudoscalar couplings.
As a result
\be
\ba{l}
\label{limfeq}
f_{e\,A}\, =\, f_{d\,A}=f_{s\,A}=f_{b\,A}\ ,
\vspace{2mm}\\
\hspace{1.5cm}
\hbox{\it get constrained by} \ K^+ \ \hbox{\it and} 
\ \Upsilon\  \hbox{\it decays}\ \ .
\ea
\ee
We get in particular, from $\,\Upsilon$ decays,
\be
\label{limfeaupsilon}
\frac{f_{e\,A}^{\ 2}}{m_U^{\,2}}\ =\ \frac{f_{b\,A}^{\ 2}}{m_U^{\,2}}\ \,\lsim\ \,\frac{G_F}{10}\ \ .
\ee

When combined with the corresponding constraint from $\,\psi\to\gamma\,U\,$ decays
this implies, for a $U$ boson having non-vanishing axial couplings, 
that the $SU(2)\times U(1)\,\times$ extra-$U(1)\,$ gauge symmetry cannot be broken down to 
$\,U(1)_{\rm QED}\,$ through the v.e.v.'s of two electroweak Higgs doublets only.
{\it An extra Higgs singlet} should acquire a (possibly large) v.e.v., in addition to the 
usual Higgs doublet v.e.v's, 
to make such effects of the longitudinal polarisation state of the $U$ boson sufficiently small, just as for the axion.

\vspace{3mm}

\noindent
{\bf \small   Consequences on the size of the cross section:}

\vspace{1.5mm}
We are interested in the possibility of producing a real $U$ boson somewhat heavier than the electron,
in $\,e^+e^-\to\gamma\ U\,$.
Disregarding for simplicity $m_e$ with respect to the energy $E$ of an incoming electron
or positron, and to $\,m_U\,$,
we get a cross section roughly proportional to
\be
\sigma\,(e^+e^-\to\gamma \,U)\ \propto\ f_{e\,V}^{\ 2}+f_{e\,A}^{\ 2}\ \ .
\ee
Not surprisingly, vector couplings of the $U$ are much less constrained (see e.g. \cite{fayetpsi})\, than axial ones (cf. eq.\,(\ref{limfeaupsilon})). {\it \,Vector couplings may well turn out to be larger}, 
\,then providing the essential contribution to the light dark matter (LDM) annihilation cross section into $\,e^+e^-\,$ pairs through the virtual production of an intermediate $U$ boson, 
also roughly proportional to 
$\,f_{e\,V}^{\ 2}+f_{e\,A}^{\ 2}\,$
\cite{boehmfayet,fermion}. They thus represent, perhaps, the best hope for 
a significant $\ e^+e^-\to\,\gamma \,U\,$ production cross section.

\vspace{2mm}

If however vector and axial couplings were related, 
as e.g. if the $U$ couplings were {\it \,chiral} 
so that $\,|f_{e\,V}|=|f_{e\,A}|$,
\be
\sigma\,(e^+e^-\to\gamma \,U)\ \,
\propto\ \,f_{e\,V}^{\ 2}+f_{e\,A}^{\ 2}\ =\ 2\ f_{e\,A}^{\ 2}\ \ ,
\ee
the strong constraints on axial couplings from (\ref{limfmua}-\ref{limfeaupsilon}) would also apply
to vector couplings, reducing significantly the hopes of detecting $U$ bosons through $\ e^+e^-\to\,\gamma \,U\,$.

\vspace{2mm}

It is thus crucial to pay a special attention to these axial couplings of the $U$. They are 
necessarily present if this one couples differently 
to left-handed and right-handed fermion fields, e.g. to $e_L$ and $e_R$.
The resulting axial coupling to the electron,

\vspace{-8mm}

\be
f_{e\,A}\ =\ \frac{f_{eL}- f_{eR}}{2}\ \ ,
\ee
could also easily induce excessively large parity-violation effects, most notably  in atomic physics,
proportional to the product $\ f_{e\,A}\ f_{q\,V}\,$ 
(cf. Fig.~\ref{fig:pvat} in Section \ref{sec:11})):
an important constraint which cannot be ignored \cite{pvat,pvat2,pvat3}.
This would be the case, in particular, if one were to consider that the $U$ 
ought to couple to the singlet right-handed electron field $e_R$, 
but not to the electroweak doublet $\,(\nu_L,\ e_L)$, in which case
\be
f_{e\,A}\ \ =\ \ -\ \frac{1}{2}\ f_{e\,R}\ \ .
\ee

\vspace{3mm}

\noindent
{\bf \small   $Z$-$U$ mixing effects:}

\vspace{2mm}

It is also crucial to pay attention to the {\it mixing effects\,} between electroweak 
($SU(2)\times U(1)$) and extra-$U(1)$ neutral gauge bosons \cite{fayet:1980rr,equiv,U}.
If the $U$ were to couple to $e_R$ but not to $(\nu_L,\ e_L)$,
or simply as soon as it couples differently to $e_L$ and $e_R$,
it should also couple to the electroweak doublet Higgs field responsible for the electron mass
$m_e$.
This corresponds in general to a situation in which there is a (small or very small) mixing between the neutral $Z$ and $U$ bosons, induced by the v.e.v.('s) of the doublet Higgs field(s),
with (small or very small) extra-$U(1)$ gauge couplings .
The fields corresponding to the physical mass eigenstates are then expressed as
\be
\label{mixing}
\left\{\ 
\ba{ccrcc}
Z^\mu\ &=&\cos\eta\ \,Z_{\ \circ}^\mu &\!+\!& \sin\eta\ \,Z"^\mu\ \ , 
\vspace{1mm}\\
U^\mu\ &=&-\,\sin\eta\ \,Z_{\ \circ}^\mu &\!+\!& \cos\eta\ \,Z"^\mu\ \ , 
\ea \right.
\ee
in terms of the standard expression of the $Z$ field,
\be
Z_{\ \circ}^\mu \ =\ \cos\theta\ W^\mu_{\ \,3}\ -\ \sin\theta\ B^\mu\ \ ,
\ee
and of the original extra-$U(1)$ gauge field, here denoted by $Z"^\mu$.

\vspace{2mm}

This small mixing does not in general affect significantly the $Z$ current.
But the current to which the $U$ boson couples
is no longer identical to the extra-$U(1)$ current,
but picks up an extra part proportional to the usual $Z$ current
$\ J^\mu_{\ Z_\circ}=$ \linebreak 
$J^\mu_{\ \,3}-\sin^2\theta\, J^\mu_{\ \,\rm em}\,$.
The $U$ couplings to $e_L$ and $\nu_L$ are then no longer constrained to be the same. 

\vspace{2mm}

As a result
{\it \,asking for a small or vanishing coupling to $\,\nu_L$}, in view of not modifying excessively the low-energy $\nu$-$e$ scattering cross section (cf. Fig.~\ref{fig:neut} in Section \ref{sec:13}),
{\it \,does not necessitate a small or vanishing coupling to $e_L$}. 
Such a requirement would imply an approximately chiral coupling to $e_R$, \,more strongly constrained than a pure $V$ coupling, and therefore a comparatively smaller $\,e^+e^-\to\gamma\,U$ cross section.

\section{$\,U\,$ bosons \,and\, LDM annihilations}
\label{sec:5}

Let us now come to dark matter, and more specifically to the possibility of 
Light Dark Matter particles, as the $U$ boson should play a crucial role in their annihilations.

\vspace{2mm}
Indeed, while weakly-interacting massive particles must in general be rather heavy,
one may now consider 
{\it light} dark matter (LDM) particles, by
using new efficient mechanisms responsible for their annihilations, most notably into $\,e^+e^-$, as shown in 
Fig.~\ref{fig:ann}.
In the absence of such new annihilation mechanisms, the relic abundance of 
LDM particles would be far too large.

\vspace{1mm}
The $U$ boson, although very weakly coupled at least to quarks and leptons, 
can still lead to the relatively ``large'' annihilation cross sections required to get the
right relic abundance ($\Omega_{\rm dm}\simeq 22\,\%$) for the non-baryonic dark matter of the Universe;
exchanges of charged heavy (e.g. mirror) fermions could play a role too,
for \hbox{spin-0} LDM particles \cite{boehmfayet}. 
$U$-induced annihilations also allow
for a $P$-wave (or mostly $P$-wave) annihilation cross section of LDM particles into $\,e^+e^-$, 
$\,<\!\sigma_{\rm ann}\,v_{\rm rel}/c\!>_{\rm halo}\,$ now, for low-velocity halo particles, being then significantly less than at freeze-out time.
(This feature may be useful to avoid a potential danger of excessive $\gamma$-ray production \cite{bes}, depending, however, on how this production occurs and is estimated.)
A gamma ray signature from the galactic center at low
energy could then be due to a light new gauge boson \cite{boehmfayet}.

\vspace{2mm}

\begin{figure}[ht]
$$
\epsfig{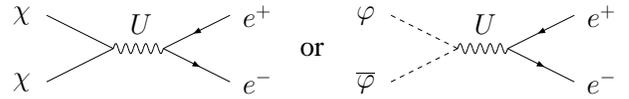}
$$
\caption{Dark matter annihilations into $e^+e^-$ pairs \cite{boehmfayet,fermion}.
The first diagram corresponds to the pair annihilation of spin-$\frac{1}{2}$ LDM particles $\,\chi$
(which may be self-conjugate, or not);
\,and the second one
to the case of spin-0 particles $\,\varphi\,$.}
\label{fig:ann}
\end{figure}

\vspace{2mm}

The subsequent observation by INTEGRAL/SPI of a bright 511 keV
$\gamma$-ray line from the galactic bulge \cite{integral}
could then be viewed as a sign of the annihilations of such positrons originating
from light dark matter annihilations \cite{boehm511}.
The annihilation cross sections of LDM particles into $e^+e^-$ are such that 
these particles, 
that could explain both the {\it \,non-baryonic dark matter\,} 
and {\it \,the 511 keV line}, may have spin $\frac{1}{2}$ instead of \hbox{spin 0} \cite{fermion}.
As of today, there is still no easy conventional interpretation for the origin of so many positrons, 
from supernovae or other astrophysical objects or processes \cite{schanne}.
The new dark matter annihilation processes mediated by $U$ exchanges, that would produce these positrons, appear as {\it stronger than weak interactions}, at lower energies
(when weak interactions are really very weak), while becoming {\it weaker than weak} 
(and therefore still difficult to detect) at higher energies.

\vspace{2mm}
The mass of the  $\,U$ boson and its couplings 
to leptons and quarks are already strongly constrained,
independently of dark matter.
Additional constraints from cosmology and astrophysics 
involve the characteristics of the LDM particles, that we shall generically call $\chi$
(irrespectively of their possible spin),
should the $U$ be responsible for their annihilations.
The main requirements are:

\vspace{2mm}
{\bf i)} \ the { \it total\,} LDM annihilation cross section 
at freeze out should be $\simeq 4$ or 5 pb, 
to get the right relic abundance, or, more precisely \cite{fermion}:
\be
\label{sigmaF}
<\!\sigma_{\rm ann}\,v_{\rm rel}/c\!>_F\, \simeq 4\ \hbox{to 5 pb}
\left\{\!\ba{l}
\small \times \ 2\ \  \hbox{if LDM not self-conjugate,}
\vspace{2mm}\\
\small \times \ \frac{1}{2} \  \hbox{if} \ S\ \hbox{instead of $P$-wave ann.}
\ea\right.
\ee

\vspace{2.5mm}

{\bf ii)} \ constraints from the intensity of the 511 keV $\gamma$-ray line from the galactic bulge involve the {\it \,partial\,} annihilation cross section for $\,\chi\,\chi\to e^+e^-\,$
at low halo velocities, and depend on whether it is $S$-wave or $P$-wave-dominated (with $\,\sigma_{\rm ann}\,v_{\rm rel}\,\propto 1\,$ or $\,v^2$, \,respectively). 
They are also sensitive to the shape of the
dark matter profiles adopted within the bulge, a $P$-wave cross section requiring a more peaked halo density \cite{asc,rasera}.

\vspace{1mm}
A $S$-wave cross section, such that 
$<\!\sigma_{\chi\chi\to e^+e^-}v_{\rm rel}/c>_{\rm halo}$ $\approx\, <\!\sigma_{\chi\chi\to e^+e^-}\,v_{\rm rel}/c>_F \ \approx\,$ 
1 to a few pb \,(given that we are dealing here with the {\it \,partial\,} annihilation cross section into $\,e^+e^-$, excluding neutrinos)
\footnote{This cross section should be $\simeq 2$ pb (doubled in the non-self-conjugate case),
times the branching ratio $B_{\rm ann}^{ee}$ for producing $e^+e^-$ 
in LDM annihilations, here assumed to be not too small. 
This $B_{\rm ann}^{ee}$ could be e.g. $\simeq 40$\% if all decay channels into
$\,e^+e^-$ or $\,\nu\bar\nu\,$ pairs contribute equally;
it could also approach 1, as $\,\nu\bar\nu$ modes may well be suppressed.},
\,would necessitate a (relatively) heavier LDM particle, say
 $\gsim 30$ MeV (as the LDM number density scales as $ 1/m_\chi$ and the 511 keV 
emissivity as $1/m_\chi^2$), which is probably excluded
as we shall see.
\vspace{1mm}

A $P$-wave cross-section, for which $\,<\!\sigma\,v_{\rm rel}\!>_{\rm halo}\,$ would be much smaller, 
would require, to get the observed 511 keV signal, a much lighter LDM particle 
( $\simeq\frac{1}{2}$ to typically a few MeV), 
with a rather peaked halo profile~\cite{rasera} (cf. Fig.~7 in that paper)\,\footnote{The profile should be sufficiently steep near the Galactic Center, e.g. a ``Moore-type'' distribution 
with $\rho \approx r^{-\gamma}$ not so far from $r^{-1.5}$, \,near the
Galactic Center.}, or a more clumpy one, in which case the mass of the LDM particle
could be higher.
Intermediate situations are also possible for a wide range of LDM masses, 
with a cross-section (\ref{sigmaF}) \,\hbox{$P$-wave} dominated at freeze-out, 
later becoming smaller and ultimately $S$-wave dominated (or $\,S+P$-wave) 
for low-velocity halo particles \cite{asc,rasera}
\footnote{In particular, we may have $S$-wave-dominated halo annihilations,
with typical $m_\chi\approx 3$ to \,30\, MeV, and
a cross section (scaling like $\,m_\chi^{\,2}$) which depends on the dark matter profile:
\vspace{-1.5mm}
$$
\vspace{-1mm}
\ \ \ \ \ \ (\sigma_{\chi\chi\to e^+e^-}v_{\rm rel})_{\rm halo} \approx(.2\ \hbox{fb}\ \, \hbox{to}\ \, .2\ \hbox{pb})\ 
\hbox{\small $\displaystyle \left(\,m_\chi/(10\ \hbox{MeV})\,\right)^2$},\!\!\!
$$
as can be seen from Fig.~7 of \cite{rasera}, small compared to the cross section at
freeze-out,
to be provided by the $P$-wave term.}.

\vspace{1mm}

Other constraints \,{\bf iii)} \,require that the LDM mass $m_\chi$ 
be sufficiently small ($\,\lsim 3$ up to maybe 30 MeV depending on the hypotheses made),
to avoid excessive $\gamma$-rays
from inner-bremsstrahlung, bremsstrahlung, and in-flight annihilations~\cite{fermion,beacom}. 
Constraints \,{\bf iv)} \,from core-collapse supernovae require LDM particles to be 
$\gsim 10$ MeV at least, if they have relatively ``strong'' interactions with neutrinos,
as they do with electrons~\cite{fhs} \footnote{In the case of a LDM particle lighter than about 10 MeV, this 
supernovae analysis points in the direction of a small coupling $\,f_\nu$
of the $U$ boson to neutrinos, significantly smaller than its coupling $\,f_e$ to electrons.
This would make it even more difficult than indicated in \cite{hoop} to attempt detecting $U$ bosons with high-energy neutrino telescopes. Indeed, having $\,f_\nu \approx f_e\, $ (then $\lsim\, 3\ 10^{-6}\ m_U$(MeV)\,, cf. Section \ref{sec:13}) 
\,would lead to relatively large neutrino-LDM interactions (as for the large LDM interactions with electrons, responsible for their annihilations into $e^+e^-$), in conflict with the results of \cite{fhs} for such light Dark Matter particles.}.
No further constraints are obtained from the evaluation of the soft $\gamma$-ray extragalactic background that may be generated by the cumulated effects of LDM annihilations, 
once one takes into account that positrons cannot annihilate in small mass halos
\cite{rasera}.

\section{ $\,e^+e^-\to\,\gamma\,U\ \, $ cross section}

\label{sec:6}

$U$ bosons may be directly produced in an accelerator experiment,
through the process $\,e^+e^- \to \gamma\ U\,$, \,as shown in Fig.~\ref{fig:eegammau}
\cite{fayetmez,boehmfayet,boro}.
This was first evaluated at threshold
($\sqrt s\simeq 2\,m_e$) long ago, assuming $\,m_U<2\,m_e$,
to discuss the production of a very light $U$  (less than about 1 MeV) remaining invisible, 
in positronium decays \cite{fayetmez}.
The relevant parameters are the mass $m_U$, and the vector and axial couplings to the \linebreak electron
appearing in the lagrangian density 
\be
{\cal L}\ =\ -\ \,U_\mu \,\ \bar e\ \gamma^\mu\ (f_{eV}-\,f_{eA}\,\gamma_5) \ \,e\,+\  ... \ \ .
\ee
These are expressed in terms of chiral couplings as 
\be
f_{e\,V}\ =\ \frac{f_{eL}+f_{eR}}{2}\ \ ,\ \ \ f_{e\,A}\ =\ \frac{f_{eL}- f_{eR}}{2}\ \ ,
\ee
${\cal P}_L=\frac{1-\gamma_5}{2}\,$ 
and $\,{\cal P}_R=\frac{1+\gamma_5}{2}\,$ denoting the left-handed and right-handed projectors, respectively.

\begin{figure}[ht]
$$
\epsfig{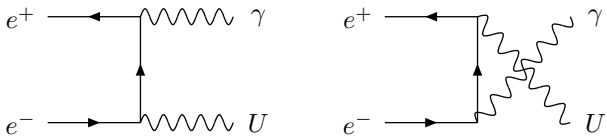}
$$
\caption{Direct production of a $\,U$ boson in $\,e^+e^-$ annihilation.
The $U$ should decay preferentially into LDM particles if $\,m_U>\,2 \,m_\chi$, 
and otherwise into $\,e^+e^-$ or possibly $\,\nu\bar\nu$ pairs.}
\label{fig:eegammau}
\end{figure}

\vspace{2mm}

We shall be interested here in the production of a $U$ heavier than 1 MeV, in $\,e^+e^-$ annihilations.
For a vector coupling of the $U$ and at high energy $\,2E\,$ large compared to $\,2\,m_e\,$ and $m_U$
so that both $m_e$ and $m_U$ may be disregarded,  the cross-section is equal to
$\,2\,f_{e\,V}^{\,2}/e^2\,$ times the $e^+e^-\to\,\gamma\,\gamma\,$ cross section.
If there is an axial coupling $\,f_{e\,A}\,$ as well, this ratio is to be replaced 
(again disregarding the electron mass $\,m_e\,$ as compared to $\,E$) by
\be
2\ \ \frac{f_{eV}^{\, 2}+f_{e\,A}^{\, 2}}{e^2}\ \ =\ \ 
\frac{f_{eL}^{\, 2}+f_{eR}^{\, 2}}{e^2}\ \ ,
\ee
also denoted $\ 2\ \,\frac{f_e^2}{e^2}\,$.
The detectability of this process depends essentially on the values of the $U$ couplings to the electron, as compared to the positron charge $\,e=\sqrt{4\pi\alpha}\simeq\,.3$.

\vspace{1mm}

At energy $\,2\,E\,$ large compared to both $m_U$ and $2\,m_e$, 
one has 
\footnote{There are corrections when the non-vanishing $m_e$ is taken into account. 
In particular, for a light $U$ an axial coupling generates 
an effective pseudoscalar coupling $\,f_{e A}\frac{2 m_e}{m_U}$;
the resulting terms in $\sigma$, proportional to $\,e^2\ (f_{eA}\ \frac{2\,m_e}{m_U})^{2}\,/s$,
can be neglected for $m_U \gsim$ a few MeV,
as compared to the $\ \approx e^2 \ (f_{eA}^{\,2}+f_{eV}^{\,2})\,/s\ $ terms in (\ref{sigmagammau}).
\ For $\,m_U< 2\,m_e\,$ these terms become essential in the evaluation
of the annihilation cross section, or positronium decay rate, 
for $\,e^+e^-\to\,\gamma \,U$ \cite{fayetmez}.}
\be
d\sigma\ (e^+e^-\to \gamma\,U)\ \simeq\ \frac{f_{eL}^{\,2}+f_{eR}^{\,2}}{e^2}\ \ 
d\sigma\ (e^+e^-\to \gamma\,\gamma)\ \ .
\ee
As
$\,
\frac{d\sigma}{d\cos\theta }\ (e^+e^-\to \gamma\,\gamma)\, \simeq \, \frac{4\pi\,\alpha^2}{s}\ (\,\frac{1}{\sin^2\theta}-\frac{1}{2}\,)\,,
$
\,and
\,$\theta$ (the polar angle of the photon produced with respect to the direction of the incoming electron) being here in the $\,[0,\pi]$ instead of $\,[0,\pi/2]\,$ interval, one has
\bea
\nonumber
\label{sigmagammau}
\frac{d\sigma}{d\cos\theta}\,(e^+e^-\!\to \gamma\,U)\ \simeq \ \frac{\alpha\ 
(f_{eL}^{\,2}+f_{eR}^{\,2})}{2\,s}\ \
\hbox{\Large$\left(\right.$}\frac{1}{\sin^2\theta}\,-\,\frac{1}{2}\hbox{\Large$\left.\right)$}\ .
\eea
\vspace{-6mm}
\be
\ee

If the $U$ mass cannot be neglected as compared to the total energy $2\,E$ of the scattering 
electrons and positrons, the cross section may be obtained from the corresponding expression for $\,e^+e^-\to\,\gamma\,Z\ $ \cite {bohmsack,boro}. This gives, neglecting again $m_e$ for simplicity\,\footnote{With
$\ t,\ u\,\simeq\ -\,\frac{1}{2}\ (s-m_U^2)\ (1\mp\cos\theta)\,$, \,we get
\vspace{-3mm}
$$
\vspace{-3mm}
\ \ \ \ \ \ \ \ \frac{s^2+m_U^{\,4}}{2\,ut}-1=\frac{2}{(s-m_U^{\,2})^2} \left(\frac{s^2+m_U^{\,4}}{\sin^2\theta}-\frac{(s-m_U^{\,2})^2}{2}\right).
$$
},

\vspace{-2mm}
\be
\label{sigmagammaumass}
\frac{d\sigma}{d\cos\theta}\
\simeq \ \frac{\alpha\ 
(f_{eL}^{\,2}+f_{eR}^{\,2})}{2\ s^2\ (s-m_U^{\,2})}\ \,\left(\,\frac{s^2+m_U^{\,4}}{\sin^2\theta}\,-\,\frac{(s-m_U^{\,2})^2}{2}\,\right)\ ,
\ee
which reduces to (\ref{sigmagammau}), for $\,s = 4\,E^2 \gg\,m_U^{\,2}\,$.
\vspace{4mm}

The $U$ boson can then decay into $e^+e^-$, or an invisible $\nu\bar\nu$ or LDM particle pair 
(the latter being favored for $m_U>2\,m_\chi$)
\footnote{$U\to\,e^+e^-\,$ may represent only $\approx$ 40 \% of the decays, 
if all $e^+e^-$ and $\nu\bar\nu$ channels contribute equally,
the $\chi\chi$ mode being absent or kinematically forbidden.
$B_{U\to e^+e^-}$ could also be very close to 0 if $m_U>2m_\chi$, 
as the $U$ is expected 
to be more strongly coupled to LDM than to ordinary particles.
It could approach 1 if $m_U<2\,m_\chi$, with the $U$ coupling
much less to neutrinos than to electrons.}.
The crucial quantity, to discuss if a light $U$ boson could be detectable in this way, is the size of its vector and axial couplings, $\,f_{eV}$ and $f_{eA}$, to the electron.
The possibility of detecting $U$ bosons at 
current $B$-factories or at the $\phi$ factory DA$\Phi$NE, which could be sensitive to couplings $f_{eR}$ larger than $10^{-4} - 10^{-3}$  (DA$\Phi$NE)
down to $3\ 10^{-5} \,-\, 3\ 10^{-4}$ ($B$-factories), has been considered recently (the first numbers correspond to 
100\,\% invisible decay modes, the last to 100\,\% decays into $e^+e^-$) \cite{boro},
using, however, specific hypothesis whose validity may be questioned
\,-- such as a chiral coupling of the $U$ of $e_R$ only, without mixing between the $Z$ and $U$ bosons
--\,
and disregarding a number of relevant constraints, most notably the strong ones involving the 
axial coupling of the $U$ to the electron.
This has the effect of being overly optimistic as to the detectability of the $U$ boson in $e^+e^-$ scatterings, by suggesting that most of the relevant parameter space could be probed soon in this way.

\section{\vspace{1mm}
Can $\,U\,$ production \,cross section 
\,be \,constrained \,from \,LDM annihilations \,?}

\label{sec:7}

Is it possible to relate the expected size of the production cross section 
for $e^+e^-\,\to\,\gamma\,U\,$ 
with the characteristics of the LDM particle, as the exchange of a virtual $U$ 
should be responsible for the LDM annihilation cross section, 
constrained both from the relic abundance of LDM particles and intensity of the 511 keV $\gamma$-ray line\,
(cf. Fig.~\ref{fig:ann} in Section \ref{sec:5})~?
Not so easily, in fact, as the former is proportional to $\,e^2\,f_e^2\,$, and the latter to
$\,c_\chi^2\,f_e^2\,$ (cf. Fig. \ref{fig:eegammau} in Section \ref{sec:6}), 
$\,c_\chi$ denoting 
the magnitude of the $U$ coupling to the LDM particle, 
denoted by $\,\chi\,$ independently of its spin, $\,\frac{1}{2}\,$
or \,0.
Some relations were presented in \cite{boro}, which however follow mostly
from specific assumptions on the size 
of the $U$ coupling to LDM particles. It is thus necessary to discuss again the possible
size of the $U$ couplings to electrons,
taking also into account a number of aspects disregarded previously.

\vspace{2mm}

Annihilation cross sections of LDM particles into $e^+e^-$ depend on the product 
$\,c_\chi f_e$,
\,as well as on $m_U$ and $\,m_\chi$, and more precisely on

\vspace{-7mm}
\be
\frac{c_\chi\,f_{e}}{|\,m_U^{\,2}-4\,m_\chi^{\,2}\,|}\ \ m_\chi\ \ .
\ee
To obtain the correct relic density we need a total annihilation cross section 
of the order of 4 to 5 pb, as follows from (\ref{sigmaF}), i.e. an annihilation cross section into $e^+e^-$ 
of the order of 4 to 5 pb, times the branching fraction $B_{\rm ann}^{ee}$.
This requires (cf. eq.~(16) in the first paper of \cite{fermion}):
\be
\label{sizecf}
|c_\chi|\ (f_{eV}^{\,2}+f_{eA}^{\,2})^\frac{1}{2}\ \simeq 10^{-6}\ 
\frac{|m_U^{\,2}-4\,m_\chi^{\,2}|}{m_\chi\ (1.8\ {\rm MeV})}\ \ 
\left(B_{\rm ann}^{ee}\right)^{\frac{1}{2}}.
\ee

\vspace{1mm}
For $m_U \simeq 10$  MeV 
and $ m_\chi \simeq 4 $ MeV as considered in \cite{boehmfayet}, or 6 MeV,
this would give 
\be
\label{limitesurcf}
|\,c_\chi\,f_e\,|\  \simeq\ 5\ 10^{-6}\ \ ,
\ee 
or $\,\simeq 3\ 10^{-6}$ only if 40\% of annihilations led to $e^+e^-$,
the rest of the required LDM annihilations being provided by the $\,\nu\bar\nu$ channels.
\,For a heavier $U$ we could get larger couplings, e.g. 
\be
\hbox{up to}\ \ \ |\,c_\chi\,f_e\,|\ \simeq \ \frac{10^{-2}}{2\ m_\chi(\hbox{MeV)}}\ , \ \  \hbox{for a 100 MeV}\ \ U\ \ .
\ee
\vspace{1mm}

Discussing, however, limitations on the product $\,c_\chi f_e\,$ 
does not help so much as we are primarily interested in the size of the coupling to the electron, represented by $f_e$.
Dividing $f_e$ (and $f_\nu$) by 10 while multiplying $c_\chi$ by the same factor 10\,
leaves unchanged the annihilation cross sections 
at freeze out, and nowadays in the halo.
But it has a crucial effect on the detectability of the $U$ boson by dividing 
its production cross section by 100\,!

\vspace{1mm}
This illustrates that
{\it dark matter considerations only play a secondary role} in the determination of the size of the couplings to the electron, $\,f_{eA}$ and $f_{eV}$,
once we have checked that suitable LDM annihilation cross sections can indeed be obtained,
with an appropriate coupling to the LDM particle $c_\chi \lsim 1$, 
or in any case $\sqrt{4\,\pi}\ $ if we would like 
the theory to remain perturbative\,\footnote{The light $U$ boson may have very small couplings with quarks and leptons, 
and a significantly larger one $\,c_\chi$ to LDM particles.}.
Still $m_U$ should in general better not be excessively large as compared to $\,2\,m_\chi$, otherwise the 
$U$ couplings to ordinary particles would tend to be too large if $c_\chi$ is to remain perturbative.

\vspace{2mm}
To quantify this, 
demanding $\,c_\chi<\sqrt{4\,\pi}\,$ would imply from (\ref{sizecf}) 
\be
\label{sizef}
f_e=(f_{eV}^{\,2}+f_{eA}^{\,2})^\frac{1}{2}\ \ \gsim \ \ 3\ 10^{-7}\ 
\frac{|m_U^{\,2}-4\,m_\chi^{\,2}|}{m_\chi\ (2\ {\rm MeV})}\ \ 
\left(B_{\rm ann}^{ee}\right)^{\frac{1}{2}}.
\ee

\vspace{2mm}

For $m_U\simeq 10$ MeV and  $m_\chi\simeq 4$ (or 6)\, MeV, the couplings to electrons
should then verify roughly, from (\ref{limitesurcf}), 
\be
f_e\ \gsim\ 10^{-6}\ \ , 
\ee
with an annihilation ratio into $e^+e^-$,
$\,B_{\rm ann}^{ee}$, \,taken to be of almost unity. 
I.e. {\it they could be quite small}, \,but may well 
also be significantly larger, 
$\,f_e\simeq 5 \ 10^{-4}\,$ corresponding in the above example to 
$\,c_\chi\simeq 10^{-2}$.

\vspace{2mm}

For larger values of $m_U$, e.g. $100$ MeV with $m_\chi=5$ \,(resp. 15) MeV, 
the couplings to electrons  should verify
\be
f_e \ \gsim \ 3\ 10^{-4}\ \  (\hbox{resp.}\ \  10^{-4})\ \ ,
\ee
so as to have $\,c_\chi<\sqrt{4\,\pi}\,$.
For $m_U=300$ MeV with $m_\chi=15$ MeV, 
$\,f_e \gsim 10^{-3}$.
In such cases the {\it $U$ couplings to electrons have to be relatively ``large''},
provided of course such values are still also compatible with all other constraints,
most notably from
from $\,g_e\!-\!2,\ g_\mu\!-\!2$, $\,\psi,\ \Upsilon$ and $\,K^+$ decays, parity-violation effects in atomic physics, $\,\nu$-$e$ scattering, as we shall discuss more precisely now.

\section{
$g_e-2\, $ constraints on $\,U$ couplings to $\, e$}
\label{sec:8}

Let us consider the contributions induced by the exchanges of a light $U$ boson to the anomalous magnetic moments of the charged leptons, electron and muon (see Fig.~\ref{fig:magn}) \cite{fayet:1980rr,pvat,pvat3,boehmfayet,fermion}.

\vspace{-4mm}

\begin{figure}[ht]
$$
\epsfig{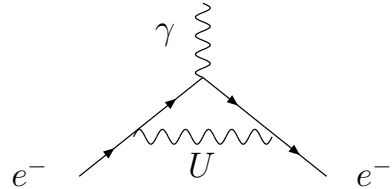}
$$
\caption{$U$-exchange contribution to $\,g_e-2$\,.}
\label{fig:magn}
\end{figure}

\vspace{-6mm}

\subsection{Vector coupling}

For a vector coupling to the electron, the additional contribution to 
$\,a_e= $ \hbox{$(g_e\!-2)/2$}\,
is given by
\be
\label{aeV}
\delta a_e^{V}\!\simeq 
\frac{f_{e V}^{\, 2}}{4\ \pi^2}\int_0^1\!\frac{m_e^{\,2}\ x^2\,(1-x)\ dx}
{m_e^{\,2}\,x^2+m_{U}^{\,2}(1-x)}
\simeq \frac{f_{eV}^{\ 2}}{12\ \pi^2}\ \frac{m_e^{\,2}}{m_U^{\,2}}\ F(\hbox{\small$\displaystyle\frac{m_U}{m_e}$}).
\ee
It would reduce to a QED-like expression $\,\frac{f_{e V}^{\, 2}}{8\ \pi^2}\,$ 
if the $U$ were much lighter than $m_e$, and to 
$\,\frac{f_{eV}^{\ 2}}{12\, \pi^2}\,\frac{m_e^{\,2}}{m_U^{\,2}}\,$
if much heavier.
For a $U$ at least as heavy as $m_e$ we tabulate $\,F(\frac{m_U}{m_e})\,$ as follows:
\bea
\nonumber
\label{tableauF}
\ \ \ \ba{|c||c|c|c|c|c|}
\hline  &&&&&\\[-1mm]
\ m_U\ & m_e & \,2\, m_e\, & \,5\, m_e\, & \,10\, m_e\, & \ \hbox{large}\ \\ [3mm] 
\hline &&&&&\\ [-1mm]
F(\frac{m_U}{m_e}) & \ \hbox{\small$\displaystyle \frac{\pi}{\sqrt 3} -\frac{3}{2}$}  \simeq .31\ &  .54 & .81 & .92& \simeq 1
\\ [4mm] \hline
\ea
\hspace{-1mm}
\\[1mm]
\eea

\vspace{2mm}

\normalsize
Taking into account the latest experimental measurement of the anomalous magnetic moment of the electron \cite{newg-2},
\be
a_e\ =\ (1 \,159 \ 652 \ 180.85 \,\pm .76)\ 10^{-12} \ \ ,
\ee
as well as improved QED calculations \cite{kinoshita}, implies, given the other uncertainties in the determination of $\,\alpha$,
that any extra contribution 
$\delta a_e$ 
should satisfy \footnote{Up to a recent time $\delta a_e$ was constrained to verify
$\,\delta a_e\simeq (1.24\pm .95)\, 10^{-11}$ \cite{kinoshita}, 
i.e., approximately, $ \,-\,10^{-11} \lsim\delta a_e \lsim  3\, 10^{-11}$.
The very recent measurement of the electron anomalous magnetic moment shifted the experimental value 
of $a_e$ downward by 1.7 $\sigma$, with an uncertainty nearly 6 times lower than in the past
\cite{newg-2}. We still remain, however, with the uncertainties in the best determinations 
of $\alpha$ independently of $a_e$,  so that we can now write approximately, from the comparison between
measured and ``calculated'' magnetic moments,
$$
-2\,10^{-11} \lsim\delta a_e \lsim  2\  10^{-11} \ \ .
$$
}
\be
|\delta a_e|\ \,\lsim\, \ 2\  10^{-11}\ \ .
\ee
This requires
\be
\label{limfve0}
|f_{eV}|\ \lsim \ \frac{10^{-4}}{\sqrt {F(m_U/m_e)}}\ \ m_U(\hbox{MeV})\ \ ,
\ee 
that we can simply remember as
\be
\label{limfve}
|f_{eV}|\ \lsim \ 10^{-4}\ m_U(\hbox{MeV})\ \ ,
\ee
or
\be
\frac{f_{eV}^{\ 2}}{m_U^{\ 2}}\ \,\lsim\,\ 10^3\ G_F\ \ ,
\ee
as soon as $m_U$ is larger than a few MeV's \footnote{The limit on $\,|f_{eV}|\,$ decreases 
down to about
$\,4\ 10^{-5}$ for $m_U$ much smaller than $m_e$, a situation which we are not interested in here.}.
We immediately see that this constraint is relatively weak, compared to those involving axial couplings as deduced from $\,g_\mu-2\,$, $\Upsilon$ decays and parity-violation effects in atomic physics ...

\subsection{Vector and axial couplings}

If there is also an {\it axial coupling} one gets 
$\, \delta a_e=\delta a_e^V+\delta a_e^A$,
with \cite{leveille}
\be
\delta a_e^A\ \simeq\, \ -\
\frac{f_{e A}^{\ 2}} {4\ \pi^2}\ \ \frac{m_e^{\,2}}{m_U^{\,2}}\ \
H(\hbox{$\displaystyle\frac{m_U}{m_e}$})
\ \ .
\ee
The quantity
\be
\displaystyle
H=\int_0^1\frac{2x^3+(x-x^2)(4-x)\,m_U^{\,2}/m_e^{\,2}}{x^2+(1-x)\,m_U^{\,2}/m_e^{\,2}}\ \ dx
\ \ , 
\ee
varies between $\,\simeq1\,$ for $\,m_U\,$ much smaller than $\,m_e$, and $\,\simeq 1.31$ for 
$\,m_U=m_e$, \,up to $\,\frac{5}{3}\,$ for $\,m_U\,$ much larger than $\,m_e$. One can write:
\be
\delta a_e\ \simeq \ \frac{f_{eV}^{\ 2}\ F(\frac{m_U}{m_e}) 
- 5\,f_{eA}^{\ 2}\ \,\frac{3}{5}\, H(\frac{m_U}{m_e})} 
{12\ \pi^2}\ \ \ \frac{m_e^{\,2}}{m_U^{\,2}}\ \ .
\ee
As soon as $m_U$ is larger than a few MeV's (cf. eq.~(\ref{tableauF}) and eq.~(\ref{tableHmu}) 
in Section \ref{sec:9}), 
one can use the simplified expression \cite{leveille}
\bea
\delta a_e\,&\simeq& \ \frac{f_{eV}^{\ 2}-5\,f_{eA}^{\ 2}} {12\ \pi^2}\ \ \ \frac{m_e^{\,2}}{m_U^{\,2}}
\vspace{4mm}\\
&\simeq&\ \frac{3\,f_{eL}\,f_{eR}-f_{eL}^{\ 2}-f_{eR}^{\ 2}}{12\ \pi^2}\ \ \ \frac{m_e^{\,2}}{m_U^{\,2}}
\ \ .
\eea
This implies, roughly, for $m_U\gsim$ a few MeV,
\be
|\,f_{eV}^{\ 2}-5\,f_{eA}^{\ 2}\,|\ 
\lsim\
10^{-8}\, m_U(\hbox{MeV})^{\,2}\ ,
\ee
or
\be
\frac{|\,f_{eV}^{\ 2}-5\,f_{eA}^{\ 2}\,|}{m_U^{\ 2}}\ \lsim\ 10^3\ G_F\ \ .
\ee

\vspace{2mm}
In general no limit can be obtained on $\,f_{eV}\,$ and $\,f_{eA}\,$ separately, 
due the possibility of cancellations between positive and negative contributions
to $\,\delta a_e$. More specifically, one gets as in \cite{pvat3} for a purely axial coupling, 
\be
\label{limfeamu}
|f_{eA}|\ \lsim \ 5\ 10^{-5}\  \,m_U(\hbox{MeV})\ \ .
\ee
Practically the same limit on $\,|f_{eA}|\,$ as in (\ref{limfeamu}) also apply in the case of a chiral coupling, e.g., right-handed, for which one has:
\be
\label{limfechiral}
|f_{eR}|\ \lsim \ 10^{-4}\ \,m_U(\hbox{MeV})\ \ .
\ee
These limits  scale with the $\,U$ mass, roughly like $\,m_U$
\footnote{Such limits could be alleviated in the presence of extra contributions 
to $a_e$ and $a_\mu$, as from heavy (e.g. mirror) fermions
in the case of spin-0 LDM particles \cite{boehmfayet}.
Conversely, $U$ exchanges could help making acceptable such heavy fermion contributions 
to $\,g-2$, \,if present.}.

\section{
 $g_\mu-2\,$ constraints on $\,U$ couplings to $\, e$
\vspace{1.5mm}\break $\hbox{\small (assuming lepton universality)}$}
\label{sec:9}

Additional constraints on the couplings of the $\,U$ 
to the electron may be obtained from the consideration of the {\it muon} $\,g-2$, under the hypothesis of lepton universality for the 
$U$ couplings.
The vector coupling of the $U$ might also be responsible for the somewhat 
large value of the muon $\,g-2$, as compared to standard model expectations, should 
this effect turn out to be real.

\subsection{Vector coupling}

For a $U$ with  a {\it vector coupling} 
to the muon, one has, as in (\ref{aeV}),
\be
\delta a_\mu\,\simeq \frac{f_{\mu V}^{\, 2}}{4\ \pi^2}\,\int_0^1\ \frac{m_\mu^{\,2}\ x^2\,(1-x)\ dx}
{m_\mu^{\,2}\ x^2+m_{U}^{\,2}\,(1-x)}\ \simeq
\frac{f_{\mu V}^{\, 2}}{8\ \pi^2} \ \,G\,(\hbox{$\displaystyle\frac{m_U}{m_\mu}$})\ ,
\ee
which reduces to $\,\frac{f_{\mu V}^{\, 2}}{8\ \pi^2}\,$,
in the limit of a light $U$ as compared to $m_\mu$.
If the $U$ is not sufficiently light, we tabulate the function
\be
G\,(\hbox{$\displaystyle\frac{m_U}{m_l}$})\ =\ \frac{2}{3}\
 \frac{m_l^{\,2}}{m_U^{\,2}}\ F\,(\hbox{$\displaystyle\frac{m_U}{m_l}$})\ \ ,
\ee 
as follows
\bea
\nonumber
\,\ba{|c||c|c|c|c|c|}
\hline  &&&&&\\[-1mm]
\ m_U\ & \hbox{small} & m_\mu/10 &  m_\mu/4 & m_\mu/2 & \ m_\mu\ \\ [3mm] 
\hline &&&&&\\ [-1mm]
\,G\,(\frac{m_U}{m_\mu})\, & \simeq   1 &  .77 & .57 & .38& 
\ \hbox{\small$\displaystyle \frac{2 \pi}{3 \sqrt 3}$} -1\, \simeq .21\
\\ [4mm] \hline
\ea
\hspace{-1mm}
\\[1mm]
\eea

The latest experimental measurement of the anomalous magnetic moment of the muon \cite{gmu-2exp},
\be
a_\mu^{\rm exp}\ \ =\ \ (11\,659\,208.0\pm 6.3) \ 10^{-10}\ \ ,
\ee
compared to improved Standard Model expectations \cite{gmu-2th},
\be
a_\mu^{\rm SM}\ \ =\ \ (11\,659\,180.4\pm 5.1) \ 10^{-10}\ \ ,
\ee
3.4 ``$\sigma$'' below the experimental value,
implies that an extra contribution to $\,a_\mu$ should satisfy 
\be
\delta a_\mu\,=\,a_\mu^{\rm exp}-a_\mu^{\rm SM}\ =\ (27.6 \pm 8.1)\ 10^{-10}\ \ ,
\ee
or  $\,(27.5 \pm 8.4)\ 10^{-10}$ according to \cite{davier}.

\vspace{2mm}

If this is considered as the sign of a real discrepancy with the Standard Model, 
it could be taken as an indication 
for the existence of a new spin-1 $U$ boson, with a vector coupling to the muon
\be
|f_{\mu V}|\ \approx\ \frac{5\ 10^{-4}}{\sqrt {G\,(\frac{m_U}{m_\mu})}}\ \ ,
\ee
i.e. $\approx (.5\ \hbox{to} \ 1)\ 10^{-3}$, for a $U$ mass of up to $m_\mu$.

\vspace{2mm}

Otherwise, we may conservatively interpret this result as indicating that
\be
\label{limiteamu}
-\,10^{-9}\ \lsim\ \delta a_\mu\ \lsim \ 5\ 10^{-9}\ \ ,
\ee
which would only imply, for a pure vector coupling of the 
$U$ to the muon,
\be
|f_{\mu V}|\ \lsim\ \ \frac{6\ 10^{-4}}{\sqrt {G\,(\frac{m_U}{m_\mu})}}\ \ ,
\ee
i.e.
\be
\label{limitefvmu}
|f_{\mu V}|\ \lsim\ \ (.6\ \,\hbox{to}\ \,1.3)\ 10^{-3}\ \ ,
\ee
for $m_U<m_\mu$.
In the natural case of a universal coupling to charged leptons, 
this limit is more constraining than (\ref{limfve}),
\,for $m_U \gsim \,7$  MeV.

\subsection{Vector and axial couplings}

If the coupling has also an {\it axial part}, we can write, as for the electron,
\be
\delta a_\mu\ \,\simeq\, \ \frac{f_{\mu V}^{\ 2}}{8\ \pi^2}\ G(\hbox{$\displaystyle\frac{m_U}{m_\mu}$})\ -\
\frac{f_{\mu A}^{\ 2}} {4\ \pi^2}\ \frac{m_\mu^{\,2}}{m_U^{\,2}}\ 
H(\hbox{$\displaystyle\frac{m_U}{m_\mu}$})
\ \ ,
\ee
with 
\be
\displaystyle
H=\int_0^1\frac{2x^3+(x-x^2)(4-x)\,m_U^{\,2}/m_\mu^{\,2}}{x^2+(1-x)\,m_U^{\,2}/m_\mu^{\,2}}\ \ dx
\ \ ,
\ee
tabulated as follows:

\vspace*{-2mm}

\bea
\label{tableHmu}
\nonumber
\ba{|c||c|c|c|c|}
\hline  &&&&\\[-1mm]
\ m_U\ & \ \hbox{small}\ & \,m_\mu/2\, & \, m_\mu\,& \ \hbox{large}\ \\ [3mm] 
\hline &&&&\\ [-1mm]
H & \simeq   1 &  1.18 & 
\ \hbox{\small$\displaystyle \frac{\pi}{\sqrt 3} - \frac{1}{2}$}\, \simeq 1.31\ & \to 
\hbox{\small$\displaystyle \frac{5}{3}$}\ 
\\ [4mm] \hline
\ea
\hspace{-1mm}
\\[1mm]
\eea

\noindent
{\it Axionlike behavior of a light $U$, for $\ m_U<m_\mu\,$:}

\vspace{2mm}

The axial contribution to the anomalous magnetic moment is superficially singular
in the limit of small $m_U$ (compared to $m_\mu$), 
which originates from expression (\ref{propagator}) of the propagator of the massive spin-1 $U$ boson, 
when its couplings involve (apparently non-conserved) axial currents.
The resulting expression of the axial current contribution gets enhanced
by a factor $\,{m_\mu^{\ 2}}/{m_U^{\ 2}}$.

\vspace{2mm}
The singularity is only apparent, as one can consider the limit in which both the mass and the couplings are small, their ratios being fixed by the extra-$U(1)$ symmetry breaking scale, 
as discussed in Sections \ref{sec:2}  and \ref{sec:4} \cite{fayet:1980rr,pvat}.
In this limit of small $m_U$ as compared to $\,m_\mu$ $\,H\to 1$, and 
the axial contribution is neither singular (even if $m_U \to 0$), nor does it disappear 
(even in the limit of small axial gauge coupling $\,f_{\mu A}$).
It has, instead, {\it a finite limit}.

\begin{figure}[ht]
$$
\epsfig{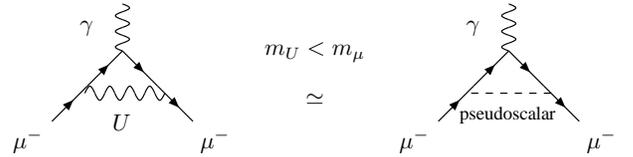}
$$
\caption{For a light $U$ as compared to $m_\mu$, \,the axial $U$-current contribution to $\,g_\mu-2$ becomes equivalent to the one due to the exchange of a quasimassless pseudoscalar
$a$, with axionlike couplings
(cf. Sections \ref{sec:2}  and \ref{sec:4}).}
\label{fig:limite}
\end{figure}

\vspace{1mm}

The axial current contribution to $\,a_\mu$ may then be written as in 
(\ref{amuaxial}-\ref{amuaxial2}),
in which $f_{\mu\,p}$, \,given by eq.~(\ref{fqlp}), denotes the effective pseudoscalar coupling of the Goldstone boson $a$ eaten away by the light $U$ \cite{fayet:1980rr,pvat}.
With $f_{\mu \,p}=2^{1/4}\ G_F^{1/2}\,m_\mu\ 1/x\,$, 
\,one recovers the
contribution of a standard axion ($A$) to \linebreak $g_\mu-2$.

\vspace{2mm}

However, the spontaneous breaking of the $SU(2)\times U(1)\ \times $ extra-$U(1)$ symmetry 
may well be due to the v.e.v's of the two Higgs doublets $\,h_1\,$ and $\,h_2$ together with an extra singlet, 
which may acquire a large v.e.v. so the the extra-$U(1)$ symmetry will then be broken at a high scale proportional to this large singlet v.e.v.. One then gets as in (\ref{couplagep2d}), taking into account $Z$-$U$ mixing effects
\cite{fayet:1980rr,equiv,U}, 
$\,f_{\mu \,p}=2^{1/4}\ G_F^{1/2}\,m_\mu\ r/x\,$, and the contribution has the same expression as for an ``invisible'' axion.

\vspace{2mm}

More precisely one has
\be
\delta a_\mu^{A}\ =\ -\ 
\underbrace{\ \frac{G_F\ m_\mu^2}{8\ \pi^2\ \sqrt 2}\ }_{\small \displaystyle \simeq 1.17\ 10^{-9}} \,H(\frac{m_U}{m_\mu})\ \ \frac{r^2}{x^2}\ \ ,
\ee
with $\ r^2/x^2 = \cos^2\zeta\ \tan^2\beta$,
\,so that expression (\ref{amuaxial2}) of $\,\delta a_\mu^A\,$ remains approximately valid as long as 
$m_U$ is smaller than $\,m_\mu$\, 
(so that $H(\frac{m_U}{m_\mu}) \lsim 1.3$). 
It also applies, approximately, when $a$ is a massive spin-0 pseudoscalar 
associated with an approximate, but explicitly broken, extra-$U(1)$ symmetry.

\vspace{2mm}
Owing to (\ref{limiteamu}), a purely axial coupling would have to verify
$r/x=\cos\zeta\ \tan\beta\lsim\,1\,$  (slightly more constraining than the $\,\lsim 1.5\,$ of \cite{fermion}), 
\,and therefore
\be
\label{limitefmua}
|f_{\mu A}|\ \,\lsim\ \ 2\ 10^{-6}\ m_U(\hbox{MeV})\ \ ,
\ee
also expressed as 
\be
\frac{f_{\mu A}^2}{m_U^{\ 2}}\ \ \lsim\ \  \frac{G_F}{3}\ \ .
\ee
Similarly
we would get for a chiral coupling, e.g.~right-handed,
\be
\label{limitefmur}
|f_{\mu R}|\ \,\lsim\,\ 4\ 10^{-6}\ m_U(\hbox{MeV})\ \ ,
\ee
approximately.
\vspace{3mm}

{\it These limits on axial couplings from $\,g_\mu-2\,$ are more restrictive than 
the corresponding ones\,}
(\ref{limfeamu},\ref{limfechiral})  {\it \ from $\,g_e-2$, 
\,by a factor $\,\approx \,25\,$.}

\vspace{3mm}

Altogether taking both $\,g_e-2\,$ and $\,g_\mu-2\,$ into consideration 
and assuming lepton universality, we get the following upper limits on the 
vector or axial lepton couplings of a $U$ boson
\be
\label{fl}
|f_{l V}|\ \lsim  \
\left\{ \ \ba{lc}
\ \ 10^{-4}\ m_U(\hbox{MeV})&\!\!\!\!\!\hbox{\small(2 MeV$< m_U \lsim 7\ \hbox{MeV})$}\,,\vspace{2mm}\\
\ 7\ 10^{-4} \ \ \hbox{up to}\ \ 1.3 \ \,10^{-3}& \hbox{\small$(m_U < m_\mu)$}\,,
\ea \right.
\ee
or
\be
\label{fl2}
|f_{l A}|\ \lsim\ \ 2\ 10^{-6}\ \,m_U(\hbox{MeV})\ \ ,
\ee
assuming for simplicity that only one of the two couplings is present.
We also get
\be
\label{fl3}
|f_{l R}|\ \lsim \ \ 4\ 10^{-6}\ m_U(\hbox{MeV})
\ee
in the case of a chiral $U$ coupling to $e_R$ and $\mu_R$, for example.
This in general decreases, especially for axial or chiral couplings
(decrease factor $\,\approx 600$), the
maximum production cross section for $\,e^+e^-\to\gamma\,U$,  \,compared to what could be inferred from $\,g_e-2\,$
only.

\section{Relating axial couplings of $\,U\,$ 
\vspace{1mm} \break
\hbox{to $\,e\,$ and $\,q$}} 

\label{sec:10}

We first assume here, as usual, that the same Higgs doublet 
\,(say $\varphi =(\varphi^+,\,\varphi^\circ)$ as in the standard model, 
or $h_1=(h_1^{\,\circ},\,h_1^-)$ as in its supersymmetric extensions)\, 
generates through $\,<\!\varphi^\circ\!>$ or $\,<\!h_1^{\,\circ}\!>$ 
\,the down-quark and  charged-lepton masses. The corresponding trilinear 
Yukawa couplings are proportional to 
$\,(\,\varphi^\dagger\ \overline{e_R}\,e_L+\hbox{h.c.}\,)\,$, 
and $\,(\, \varphi^\dagger\ \overline{d_R}\,d_L+\hbox{h.c.}\,)\,$;
or to
$\,(\, h_1\ \overline{e_R}\,e_L+\hbox{h.c.}\,)\,$, 
and $\,(\, h_1\ \overline{d_R}\,d_L+\hbox{h.c.}\,)\,$,
$\,SU(2)$ gauge indices being omitted for simplicity.

\vspace{2mm}
The gauge invariance of these trilinear Yukawa couplings 
requires, for the gauge quantum numbers $f$ associated with $U$ boson exchanges
(cf. the general analysis in \cite{U})
\be
\label{gaugeinv}
f_{e_R}\ =\ f_{e_L}+\ f_{h_1}\ \ ,\ \ f_{d_R}\ =\  f_{d_L}+\ f_{h_1}\ \ ,
\ee 
and therefore, with
$\ f_{e\,A}=$ {\large $\frac{f_{e\,L}-f_{e\,R}}{2}$}\,, \ $f_{q\,A}=$
{\large $\frac{f_{q\,L}-f_{q\,R}}{2}\,$},
\be
\label{aeaq}
f_{e\,A}\ =\ f_{d\,A}\ =\ -\,\frac{1}{2}\ f_{h_1}\ \
\ee
(or $\,\frac{1}{2}\ f_\varphi$).
\,The axial coupling of the $\,U$ 
to the charge $\,-\,\frac{1}{3}\,$ $\,d, \ s$ or $b$ quarks,
fixed by the $U$ coupling to $h_1$, 
\,should then be the same as for the $e, \ \mu$ or $\tau$ leptons:
\be
\label{faeq}
f_{e\,A}\, =\, f_{\mu\,A}\, =\, f_{\tau\,A}\ \,=\,\ f_{d\,A}\, =\, f_{s\,A}\, =\, f_{b\,A}\ \ .
\ee

\vspace{1mm}

We also get, in a similar way,
\be
\label{fau}
f_{u\,A}\, =\, f_{c\,A}\, =\, f_{t\,A}\ \ \ =\ -\,\frac{1}{2}\ f_{h_2}\ \ 
\ee
(or $\,-\frac{1}{2}\ f_\varphi$),
\,but this will not be of direct interest to us here.

\vspace{2mm}

This takes into account possible mixings between $Z$ and $U$ gauge bosons,
as we wrote  eqs.~(\ref{gaugeinv}) directly in terms of the $U$ gauge couplings, 
rather than considering the extra-$U(1)$ gauge quantum numbers $F$ in an intermediate step, 
then mixing the corresponding extra-$U(1)$ current with the standard $Z$ current $\,J^\mu_{\ 3}-\sin^2\theta\,J^\mu_{\ \rm em}\,$ to get the $U$ current.

\vspace{2mm}
Indeed eqs.\,(\ref{gaugeinv}-\ref{faeq}) may be applied as well, both to the extra-$U(1)$ gauge quantum number $F$, and to the couplings of the standard electroweak neutral gauge field
$\,Z^\mu_{\ \circ}=\cos\theta \,W_{\ \,3}^\mu-\sin\theta\,B^\mu\,$ 
to the usual weak neutral current
$\,J_{\ Z_\circ}^\mu=J_{\ 3}^\mu-\sin^2\theta\,J_{\rm \,em}^\mu\,$.
The axial part of this current, $\,J_{\ Z_\circ\, \rm ax}^\mu\equiv J_{\ 3\, \rm ax}^\mu\,$, 
\,satisfies eqs.\,(\ref{aeaq},\ref{faeq}) as well as eq.\,(\ref{fau}).

\vspace{2mm}

The conclusions (\ref{faeq}) on the universality of the axial couplings of the down quarks
and charged leptons \,-- and similarly, (\ref{fau}) for up quarks --\, remain valid even if several Higgs doublets 
are responsible for
the charged-lepton and down-quark masses, on one hand, and up-quark masses, on the other hand,  
as long as they all have the same gauge quantum numbers as $h_1$  and $h_2$, respectively.

\vspace{2mm}
Therefore as soon as we get interested in a situation involving {\it axial couplings to the electron}, or muon,
it is necessary to consider {\it axial couplings to the quarks as well}.
Strong constraints on $f_{q\,A}$ from $\psi,\ \Upsilon$ or kaon decays (cf. 
Section.~\ref{sec:11}) may then be turned into strong constraints on $f_{e\,A}$. 
All this goes in the direction of a more restrictive parameter space, leaving less room open
for an easy detection of a light $U$ boson in $e^+e^-$ scattering experiments.

\vspace{3mm}

\noindent
{\it Further implications in case of a chiral coupling to electrons:}

\vspace{2mm}

If in addition we were to decide that the $U$ is coupled to $e_R$ only, not to $e_L$, these very strong constraints on $f_{q\,A}$ and therefore on $f_{e\,A}$ would apply to the vector coupling 
to the electron $f_{e\,V}= \,-\,f_{e\,A}$ as well.
These constraints \,-- as compared to those coming from $g_e-2$ --\,
tend to diminish significantly by the maximum possible size of the $U$ coupling to the electron 
by a factor $\,\approx 50\,$.
I.e. typically from the $\,|f_{e\,R}| \lsim 10^{-4} \ m_U$(MeV) of \,(\ref{limfechiral}) 
corresponding to  $\,|f_{e\,A}| \lsim 5\ 10^{-5} \ m_U$(MeV) of \,(\ref{limfeamu}),
\,down to $\,|f_{e\,A}| \lsim  10^{-6} \ m_U$(MeV)\,. 
\,Given that in this case $\, f_{e\,V}^{\,2}\!= f_{e\,A}^{\,2}=\frac{1}{2}\ f_e^{\,2}=\frac{1}{4}\ f_{e\,R}^{\,2}\,$,
\,this corresponds roughly to
\be
\ba{r} \displaystyle
\hbox{\small maximum}\ \, \frac{f_e^{\,2}}{m_U^{\ 2}}\ \ \hbox{{\it decreased}\ \ 
from\,  }\ \approx\ \,  500\ \, G_F\ \ \hbox{from}\ \ g_e-2
\vspace{2mm}\\
\hbox{down to}\ \ \approx\ \hbox{\small $\displaystyle\frac{G_F}{5}$}\ \ \ \hbox{from} \ \ \Upsilon\ \ \hbox{decays}\ .
\vspace{2mm}\\
\ea
\ee

\vspace{2mm}

\noindent
The resulting possible $U$ production cross section in $\,e^+e^-$ annihilations 
would then be decreased {\it by more than 3 orders of magnitude}, as compared to what an optimistic
but excessively crude analysis could have indicated in such as case.
This could ruin, or in any case severely impede, the chances of finding the $U$ boson directly
in this way, in the near future.

\section{\vspace{1mm}
Restrictions \,on \,axial \,coupling \,to $\, e$ \hbox{from \,quark couplings}}

\label{sec:11}

The easiest way through which a $U$ boson could manifest, and in general 
be quickly excluded, would be through {\it flavor-changing neutral current} processes.
Fortunately in the simplest cases its couplings to quarks are found
to be flavor-conserving, as a consequence of the extra-$U(1)$ gauge symmetry
of the (trilinear) Yukawa interactions responsible for quark and lepton masses,
which naturally avoids prohibitive FCNC processes \cite{equiv,U}.

\subsection{Constraints from searches for axionlike particles}
\label{subsec:10A}

Searches for unobserved axionlike particles in the decays $\,\psi\to\gamma\,U$, 
$\,\Upsilon\to\gamma\,U$, as shown in Fig.\,\ref{fig:upsilon}, with the $U$ decaying into unobserved LDM 
or $\nu\bar\nu$ pairs, strongly constrains possible axial couplings to heavy quarks.
These radiative decays of the $\psi\,$ and the $\Upsilon\,$, which are \hbox{$\,C=-\,$} states like the photon, proceed only through the {\it \,axial\,} coupling of the $U$ boson to quarks, which has $\,C=+$.

\begin{figure}[ht]
$$
\epsfig{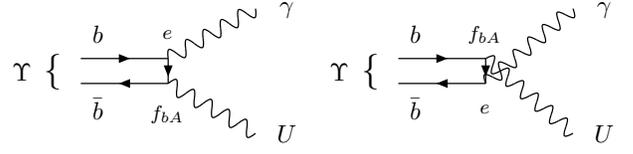}
$$
\caption{Upsilon decay $\,\Upsilon\,\to\,\gamma\,U$\,, induced by the axial coupling $\,f_{b\,A}\,$ of the $\,U$ boson to the $b$ quark. See also Fig.\,\ref{fig:ups2}.}
\label{fig:upsilon}
\end{figure}

\begin{figure}[ht]
$$
\epsfig{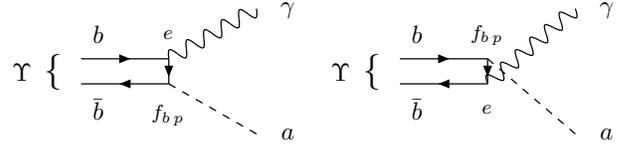}
$$
\caption{When the $U$ is light as compared to the $\Upsilon$, the sum of the decay amplitudes for $\,\Upsilon\,\to\,\gamma\,U$\,(Fig.\,\ref{fig:upsilon}) 
is essentially the same as for the production of a spin-0 pseudoscalar $a$ in
$\,\Upsilon\,\to\,\gamma\,a,$
\,with a pseudoscalar $a$ coupling to the $b$ quark  
$\,f_{b\,p}=f_{b\,A}\ \frac{2\,m_b}{m_U}\ =\ 2^{1/4}\ G_F^{1/2}\ m_b\ \frac{r}{x}
\ =\ 2^{1/4}\ G_F^{1/2}\ m_b\ \cos\zeta\ \tan\beta $\ \ \cite{fayet:1980rr,fayetmez,cras}.}
\label{fig:ups2}
\end{figure}

\vspace{2mm}
Let us also indicate that the {\it \,vector\,} coupling of the $U$ to quarks, 
which has $\,C=-$, \,can contribute,
very much as in \cite{fayetkaplan}, to the invisible decays of the $\,\psi\,$ and the $\,\Upsilon\,$ 
\be
\psi\ \ (\hbox{or}\ \ \Upsilon)\ \ \stackrel{f_{q\,V}\,c_\chi}{\longrightarrow}\ \ \chi\,\chi\ \ ,
\ee
producing a pair of two invisible LDM particles.
From the new Belle upper limit \cite{upsilon}
\be
B\,(\Upsilon\to\hbox{invisible})\ <\ 2.5\ 10^{-3}\ \ ,
\ee
we can deduce as in \cite{fayetpsi} the upper limit
\be
|\,c_\chi\,f_{b\,V}\,|\ \ <\ \ 1.4\ 10^{-2}\ \ 
\ee
for the pair-production of self-conjugate Majorana particles
(resp. $2\ 10^{-2}$ for spin-0 LDM particles, or $10^{-2}$ for Dirac particles),
\,improved by a factor $\,\sqrt {20}\,$ as compared to the earlier ones obtained from
the CLEO limit of \,5\,\%. 

\vspace{2mm}
Let us now return to the radiative decays of the $\psi$ and $\Upsilon$.
According to the analysis and evaluations of \cite{fayet:1980rr} the production rates of $U$ bosons
in these radiative decays are the same as for the equivalent (``eaten away'') pseudoscalar Goldstone boson, $a$. (The same applies if this $a$ is a massive but light pseudoscalar, associated with
a small explicit breaking of the global extra-$U(1)$ symmetry.) 
If we were working with two Higgs doublets only without introducing an extra singlet, the decay rates would be essentially the same as for a standard axion, evaluated in  \cite{wilc}.
As we also introduced an extra Higgs singlet 
which can acquire a (possibly large) v.e.v.,
the \hbox{spin-1} $U$ boson behaves like a doublet-singlet combination $a$ expressed as in (\ref{expressiondea}),
the $\,\psi$ and $\Upsilon$ decays rates being multiplied
by a factor
$\,r^2=\cos^2\zeta \,<\,1\,$.

\vspace{2mm}
The effective pseudoscalar couplings of this equivalent pseudoscalar $a$ to the $c$ and $b$ quarks 
are given by,
\be
\left\{\ 
\ba{ccl}
f_{c\,p}\ =\ f_{c\,A}\ \frac{2\,m_c}{m_U}& =&2^{1/4}\ G_F^{1/2}\ m_c\ r\ x\ 
\vspace{2mm}\\
&&\ = \ 2^{1/4}\ G_F^{1/2}\ m_c\ \cos\zeta \ \cot \beta\ \ ,
\vspace{4mm}\\
f_{b\,p}\ =\ f_{b\,A}\ \frac{2\,m_b}{m_U}& = &2^{1/4}\ G_F^{1/2}\ m_b\ r/x
\vspace{2mm}\\
&&\ = \ 2^{1/4}\ G_F^{1/2}\ m_c\ \cos\zeta \ \tan \beta\ \ ,
\vspace{2mm}\\
\ea \right.
\ee
corresponding to
\be
\label{fafp}
\left\{
\ba{lcl}
f_{c\,A}\, =\, 2^{-3/4}\ \,G_F^{1/2}\ m_U\ r\ x \!&\simeq & 2\ \,10^{-6}\ m_U\hbox{(MeV)}\ \ r\,x\  ,
\vspace{2mm}\\
f_{b\,A}\, =\, 2^{-3/4}\ \,G_F^{1/2}\ m_U\ r/x \!&
\simeq &2\ \,10^{-6}\ m_U\hbox{(MeV)}\ \ r/x\,.
\vspace{2mm}\\
\ea \right.
\ee

\vspace{1mm}

The resulting decay rates, obtained from
\be
\frac{B (\psi \to \gamma \  U/a)}{B(\psi \to \mu^+\mu^-)}\,=\, 
\frac{G_F\,m_c^2}{\sqrt 2 \pi\alpha}\ r^2\,x^2\ C_\psi
\,\simeq\ 8\ 10^{-4}\ r^2\,x^2\ C_\psi,
\ee
and
\be
\frac{B (\Upsilon \to \gamma \  U/a)}{B(\Upsilon \to \mu^+\mu^-)}\ =\ 
\frac{G_F\,m_b^2}{\sqrt 2 \pi\alpha}\ \frac{r^2}{x^2}\ C_\Upsilon
\ \simeq\ 8\ 10^{-3}\ \frac{r^2}{x^2}\ \ \ C_\Upsilon,
\ee
are
\be
\left\{   \  \begin{array}{ccll}
B \ (\ \psi \ \to \ \gamma \  \ U/a\ ) &\simeq &
  \ 5 \ \ 10^{-5} \ \ \,\ r^2 \ x^2 \  \ \ &C_{\psi} \ \ , \nonumber 
\vspace{2mm}\cr
B\  (\ \Upsilon \ \to \ \gamma\  \ U/a\ ) &\simeq &
 \  2 \ \ 10^{-4} \ \ \,(r^2/x^2) \ \  &C_{\Upsilon}   \ \ .         \cr
\end{array}   \right.
\ee
$C_{\psi}$ and $C_{\Upsilon}$, expected to be larger than $1/2$, take
into account QCD radiative and relativistic corrections. 
A $\,U\,$ boson decaying  into LDM particles 
(or $\,\nu\, \bar \nu\,$ pairs) would remain undetected.

\vspace{2mm}

From the experimental limits \cite{edwards,upsilongamma}
\begin{equation}   
\left\{   \  \begin{array}{ccl}
B\ (\ \psi \ \to \  \gamma \,+\,  \hbox{invisible}\ ) \ & <&\ \, 1.4 \ \ 10^{-5}\ \ ,  
\vspace{1mm}\cr
B\ (\ \Upsilon \ \to \ \gamma \,+\, \hbox{invisible}\ ) \  & < & \ \, 1.5 \ \ 10^{-5}
\ \ , \cr
\end{array}   \right. 
\end{equation}
we deduced 
$\ rx  < .75\,$ and $\ r/x <.4 $\ \cite{fayet:1980rr,fermion,fayetpsi,cras},
and therefore
\be
r^2\ =\ \cos^2\zeta\ <\ .3\ \ ,
\ee
which already implises that $a$ must be mostly singlet \ ($\,\sin^2\zeta >\,70$\,\%),\ 
rather than doublet \ ($\,\cos^2\zeta <\,30$\,\%).

\vspace{2mm}

This immediately implies, for the $\,\psi$, 
\,an expected branching ratio that is rather small, for example
\be
B \ (\ \psi \ \to \ \gamma \  \ U/a\ )\ \ \lsim\ 10^{-7}\ \ ,
\ee
if one is to consider also relatively large values of $\,1/x=\tan\beta=v_2/v_1\,\gsim\,10$.
Such large values of $\,\tan\beta\,$ could comparatively enhance the 
branching ratio for $\,\Upsilon\to \gamma \  U/a\,$, \,which is proportional to 
$\ \cos^2\zeta \ \,\tan^2\beta\,$.

\vspace{2mm}

These limits may be turned from (\ref{fafp}) into upper limits on the axial coupling of the $U$ to the $c$ and $b$ quarks, 
\be
\label{facb}
\left\{\ 
\ba{ccc}
|f_{cA}| &\lsim& 1.5 \ 10^{-6} \ m_U(\hbox{MeV})\ \ ,
\vspace{2mm}\\
|f_{bA}| &\lsim&  .8 \ \,10^{-6} \ m_U(\hbox{MeV})\ \ .
\ea \right.
\ee
This corresponds, approximately, to
\be
\frac{f_{b A}^{\,2}}{m_U^{\,2}}\ \ \lsim\ \,\frac{G_F}{10}\ \ .
\ee

\vspace{2mm}

By searching for the decay $\,K^+\to\pi^+ +$  invisible $U$ 
(constrained to have a branching ratio smaller than $\,\approx \,10^{-10}$
\cite{kpiu}, for $m_U<100$ MeV), which could be induced at a too large rate even 
in the absence of $s\to d\,U$ decays at tree level, with the $U$ directly attached to e.g. a $\,s$ quark line,
one may also get (see \cite{fayetpsi} for details),
\be
f_{sA}\ \lsim \ 2 \ 10^{-7} \ m_U(\hbox{MeV})\ \ .
\ee

\vspace{2mm}
Of course these limits should be somewhat relaxed for a rather light $\,U$ having a mass smaller 
than $\,2\,m_\chi$, \,and a smaller coupling to neutrinos than to electrons. The $U$ would then decay mainly into $e^+e^-$ pairs, and the size of its axial couplings to quarks would be less strongly constrained, as e.g. from \cite{mageras}, from the production of $\,\gamma\,e^+e^-$ in the final state.
This could make it desirable to get improved limits on the decays 
$\ \psi \,\to\, \gamma\ U\,, \ \ \Upsilon \,\to\, \gamma\ U\,, \ \ K^+\to\,\pi^+\,U$, 
\,with $\ U\to e^+e^-$.

\vspace{2mm}

If eqs.~(\ref{faeq}) relating the axial coupling of the electron to the axial couplings of the
\ ($d,\ s,\ b$)\ \,quarks hold, we should have, from eqs.~(\ref{faeq},\ref{facb}),
\be
\label{faelim}
f_{eA}\ \lsim  \ 10^{-6} \ m_U(\hbox{MeV})\ \ .
\ee
This upper limit on the axial coupling of the $U$ to the electron is more severe than the ones
(\ref{limfeamu},\ref{limitefmua}) 
that may be derived from the consideration of the anomalous magnetic moment of the electron, 
and of the muon assuming lepton universality.

\subsection{Constraints from \,parity-violation\,  in \,atomic physics}
\label{subsec:10B}

Experiments looking for {\it parity-violation effects in atomic physics} 
constrain the product of the axial coupling of the $U$ to the electron $f_{eA}$, times its
(average) vector coupling to a quark (cf. Fig.~\ref{fig:pvat}) \cite{pvat,pvat2} to be very small, typically
\be
\frac{|f_{eA}\,f_{qV}|}{m_U^{\ 2}}\ \lsim\,10^{-3}\ \,G_F\ \ ,
\ee
 or more precisely \cite{pvat3}:
\be
-1.5\ 10^{-14}\ m_U(\hbox{MeV})^2\, \lsim \,f_{eA}\,f_{qV}\, \lsim\, .6\ 10^{-14}m_U(\hbox{MeV})^2.
\ee

\begin{figure}[ht]
$$
\epsfig{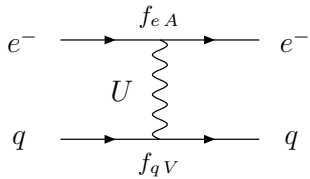}
$$
\caption{$U$-exchange amplitude contributing to parity-violation effects in atomic physics
\cite{pvat,pvat2,pvat3}.}
\label{fig:pvat}
\end{figure}

\vspace{2mm}

\noindent
These limits, valid in the local limit approximation for $m_U \geq 100$ MeV, should be multiplied by a corrective factor $K^{-1}(m_U) \geq 1$,
\,which is about 2 for  $m_U$ of a few MeV's.

\vspace{2mm}
Axial couplings to the electron that would approach a few times $10^{-5}\ m_U$(MeV), 
as considered previously (only from $g_e-2$, cf. eq.~(\ref{limfve})), would require the effective vector coupling 
to quarks to be extremely small, 
$
\,|f_{qV}|\lsim \,10^{-9} \ m_U(\hbox{MeV})\,
$.

\vspace{2mm}
Even if we were deciding to ignore the strong constraint (\ref{faelim}) from $\,\Upsilon$ decays, having
\be
|f_{eA}|\  \gsim \ 10^{-6}\ m_U\hbox{(MeV)}
\ee
would require 
\be
|f_{qV}|\ \lsim\ \hbox{a few}\ 10^{-8} \ m_U\hbox{(MeV)}\ \ ,
\ee
still very restrictive.

\vspace{3mm}

\noindent
{\it 
A $\,U$ coupled only to leptons (and dark matter),
not to quarks\,?}

\vspace{2mm}

But maybe the $U$ does not couple to quarks at all\,?
As quarks and leptons usually acquire their masses through trilinear Yukawa couplings 
to the same Higgs doublet (or doublet pair $h_1$ and $h_2$, in a supersymmetric theory), 
demanding that the extra $U(1)$ does not act on quarks
implies that it does not act on Higgs doublets either.
This leads to an extra-$U(1)$ current proportional to the leptonic current
(or to $\,L_e$, 
\,or $\,L_e -L_\mu$, \,or $\,L_e-L_\tau$, \,...), \,plus an additional dark matter contribution.
The $U$ current is here identical to this extra-$U(1)$ current 
as the extra-$U(1)$ gauge boson does not mix with the standard electroweak $Z$ boson.
But, although we would no longer have to worry about 
the strong constraints from hadronic decays, 
or parity-violation effects in atomic physics,
we still have to take into account another constraint in the leptonic sector, 
coming from the fact that $\,U$ exchanges should not modify excessively the 
neutrino-electron scattering cross section, which has been measured at low $|q^2|$ 
(cf. Sec. \ref{sec:13}).

\section{\vspace{1.2mm}  Satisfying \,constraints 
\,on \,axial \,couplings,
\,with \,a \,vectorial  $\,U$ \,current }

\label{sec:12}

\vspace{-1mm}

A simple way to satisfy automatically such stringent limits 
involving axial couplings would be to consider situations,
natural in a number of models,
in which the $U$ couples to leptons and quarks {\it in a purely vectorial} (or almost purely vectorial)
{\it \,way} \cite{equiv,U}. This is the case if there is only one Higgs doublet
(+ at least one extra singlet so that the $U$ gets its mass). 
Or several Higgs doublets (of $h_1$-type and $\,h_2$-type
as in supersymmetric theories) taken to have the same value of the extra-$U(1)$ quantum number $F$ 
once they have the same value $\,Y=-\frac{1}{2}\,$, \,or $\,+\frac{1}{2}$, of the weak hypercharge; plus again at least one extra singlet.

\vspace{2mm}

Mixing effects between the neutral $Z$ and $U$ bosons, as described by (\ref{mixing}),
in general affect the couplings of the $\,U$.
The vector part in the quark and lepton contribution to the $U$ current then normally appears
as a combination of the $B$, $L$ (or $B-L$ in a grand-unified theory, and electromagnetic currents. 
The axial part may well be absent, depending on the theory considered
(i.e. depending on the extra-$U(1)$ gauge quantum numbers chosen for the electroweak Higgs doublets).
There is also of course, in addition, an extra LDM part.

\vspace{3mm}

\noindent
{\it A vectorial $U$ current\,:}

\vspace{2mm}
The possible absence of an axial part in the $U$ current
provides a favorable situation, in view of having
 ``large'' (vectorial) couplings to electrons. 
This is also in agreement with eqs.\,(\ref{faeq}), which imply that in the absence 
of axial couplings to quarks there should be no axial coupling to leptons either.
In that case the $U$ current, purely vectorial as far as quarks and leptons are concerned, 
is expressed as a linear combination of the (conserved or almost conserved) $B$ and $L$ currents
with the electromagnetic one. With, in particular, $f_{eL}=f_{eR}$, bounded by (\ref{fl}) from the anomalous magnetic moments of charged leptons.

\noindent
\section{Consequences of a constraint \vspace{1mm}
\break 
from $\ \nu - e\ $  scattering}

\label{sec:13}

Even in such a ``favorable case'' of a vectorial coupling to quarks and leptons, 
allowing for the possibility of a larger coupling $f_e$, 
we still have to take into account  
another stringent constraint
in the purely leptonic sector, namely, from low-$|q^2|$ \,$\nu$-$e$  scattering \cite{nue},
\be
\label{fnufe}
\frac{|f_\nu\ f_{e}|}{m_U^{\,2}} \ \lsim \ G_F\ \ ,
\ee
for $m_U$ larger than a few MeV's \cite{boehmfayet}. If
we could say that the $U$ is not (or very little) coupled to neutrinos, 
this constraint would be trivially satisfied, 
and we would only have to take into account the constraints from the electron and muon $\,g-2$.

\begin{figure}[ht]
$$
\epsfig{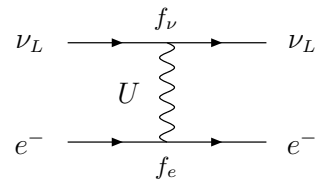}
$$
\caption{$U$-exchange amplitude, contributing to low-energy $\nu$ - electron scattering.}
\label{fig:neut}
\end{figure}

\vspace{1mm}
If we were to assume no couplings to quarks,
which results in a coupling to the leptonic currents only 
(plus a dark matter contribution),
the $U$ couplings to $e$'s and $\nu$'s, $f_e$ and $f_\nu$, should then be equal, and thus 
cannot be too large: 
\be
\label{fe}
\hbox{\underline{if}}\ \ \ f_{\nu} \, \approx\, f_{e}\ \ \ \Longrightarrow\ \ f_{e}\ \lsim\  3\ \,10^{-6}\ \,m_U(\hbox{MeV})\ \ ,
\ee
which is about $\,3\ 10^{-5}\,$ at 10 MeV, 
reducing further (compared to the $\ \approx 10^{-4}\ m_U$(MeV) of (\ref{limfve}) 
or $\ \approx 10^{-3}$ of (\ref{fl})) the hopes of detecting a light $U$ in $e^+e^-$ annihilations);  \,up to $\,\approx 10^{-3}$ at 300 MeV.

\vspace{2mm}

This upper limit (\ref{fe}) is still larger than the lower one (\ref{sizef})
from the annihilation cross section, using the requirement that the  coupling to 
the dark matter particle $\chi$ remains perturbative
(unless $\,m_U$ is taken too large as compared to $2\,m_\chi$).
The same conclusions are reached as long as we consider $f_e$ and $f_\nu$ to be
of the same order.

\vspace{2mm}

If the $U$ couplings to electrons and neutrinos turn out to be
similar, they should verify as in (\ref{fe})
$\,|f_{e}|\,\lsim \,3\ 10^{-6}\ m_U(\hbox{MeV})\,$,
\,much more constraining than the $\,\approx 10^{-4}\ m_U(\hbox{MeV})$ of (\ref{limfve}) from $\,g_e-2$. 

\vspace{2mm}

Even in this case, {\it a 100 MeV (300 MeV) $\,U$ would allow for 
a coupling to the electron of up to $\,\approx3\ 10^{-4}$ (resp. $10^{-3}$), that could be detectable},
especially if the $U$ decays invisibly into $\chi\chi$ pairs.

\vspace{3mm}

Otherwise, one may also satisfy the above leptonic constraint (\ref{fnufe})
while allowing for {\it \,couplings to electrons larger than in\,} (\ref{fe}),
by having {\it very small or even vanishing couplings to neutrinos}.
This requires taking into account mixing effects
between the $Z$ and $U$ bosons, if we want the coupling to the electron to be purely vectorial.

\section{Conclusions}

\label{sec:14}

In summary, constraints which do not involve dark matter directly,
as from an axionlike behavior of a $U$ boson (tested in $\,\psi,\ \Upsilon$ and $K^+$ decays, ... ) or atomic-physics parity-violation, as well as $Z$-$U$ mixing effects, 
cannot be ignored.

\vspace{2mm}

Constraints 
involving Dark Matter particles do not in general provide useful bounds on the expected size 
of the $U$ couplings to electrons. In particular, these couplings may well be rather small, 
provided the $U$ coupling to LDM particles be large enough, still providing 
annihilation cross sections for light dark matter particles into $e^+e^-$ 
of the appropriate size.

\vspace{2mm}

A way to satisfy systematically all strong constraints involving axial
couplings of the $U$ boson would be to consider a
$U$ coupled to a purely-vectorial neutral current, as far as quarks and 
charged leptons are concerned.
An even more favorable situation, to allow  for relatively ``large'' couplings to electrons,
is obtained when the $U$ is much less coupled to neutrinos than to electrons, 
thanks to $Z$-$U$ mixing effects
\cite{U}, as also useful to obey the supernovae constraint on lighter 
dark matter particles \cite{fhs}.

\vspace{2mm}

The $g-2$ constraints (\ref{limitefvmu},\ref{fl}) allow for 
a vectorial coupling to charged leptons of up to $\,\approx$ $ (.6$ \,to \,$1.3)\ 10^{-3}\,$ 
for $\,m_U<m_\mu$ (from $\,g_\mu-2\,$ assuming lepton universality, in the absence of any special cancellation effect). 
The constraints from \hbox{$g_\mu-2$} are then stronger than those from $g_e-2$, 
as soon as the $U$ is heavier than about 7 MeV.
In such a case, a vectorial $U$ coupling to charged leptons $\,(f_{l\,V})$ of the order of $\,10^{-3}$ 
\,could also be responsible for the 
rather large value of the muon $\,g_\mu-2$, 
\,as compared to standard model predictions, without affecting excessively the $\,g_e-2\,$ of the electron.

\vspace{2mm}

Having 

\vspace{-9mm}

\be
f_e^{\,2}\ \,\lsim\ \,10^{-6}\ \ ,
\ee
i.e. $\lsim 10^{-5}\ e^2\,$, or 
$\,f_e^{\,2}/(4\,\pi)\lsim 10^{-7}$,
\,makes in any case the detection of $U$ production in $e^+e^-$ colliders 
difficult. It is even more so if the $U$ current has vector and axial parts of comparable magnitudes,
axial couplings being very strongly constrained.
The prospects for actually producing and detecting such very weakly coupled $\,U$ bosons in $\,e^+e^-\to\,\gamma\,U$
appear as challenging, and efforts should be pursued in this direction.

\vspace{1mm}
It may also be worth considering situations in which a light spin-1 $U$ boson is produced, for example in $e^+e^-$ scatterings, through an axial coupling to the muon, $\tau$, or a heavy quark
(as we saw for $\psi$ and $\,\Upsilon$ decays), especially the $b$ (owing also to the $\,\tan\beta\,$ in its effective coupling).
The corresponding effective pseudoscalar couplings, enhanced by factors $\,2\,m_{q,l}/m_U\,$, 
\,are given by (\ref{couplagep2u}-\ref{faxr}), 
\,as for a relatively light neutral pseudoscalar Higgs boson, in supersymmetric extensions 
of the Standard Model.

\end{document}